\UseRawInputEncoding
\documentclass[superscriptaddress, twocolumn, amsmath, amssymb, aps,pra, notitlepage,longbibliography]{revtex4-1} %preview twocolumn
\usepackage{graphicx,graphics,epsfig,subfigure,times,bm,bbm,amssymb,amsmath,amsfonts,amsthm,mathrsfs,MnSymbol,physics}
\usepackage[matrix,frame,arrow]{xypic}
\usepackage[normalem]{ulem}
\usepackage{slashed}
\usepackage{dcolumn}
\usepackage{tabularx}
\usepackage{amsopn}
\usepackage{color}
\usepackage[usenames,dvipsnames,svgnames,table]{xcolor}
\usepackage[english]{babel}
\usepackage{verbatim}
\definecolor{darkblue}{rgb}{0.0,0.0,0.3}
\usepackage[colorlinks=true,
            linkcolor=red,
            urlcolor= darkblue,
            citecolor=blue]{hyperref}

\newcommand{\bea}{\begin{eqnarray}}
\newcommand{\eea}{\end{eqnarray}}

\begin{document}

\title{Quantum Mpemba effect in parity-time symmetric systems}
\author{Wanchen Ma}
\affiliation{Department of Physics, Institute for Quantum Science and Technology, Shanghai Key Laboratory of High Temperature Superconductors, International Center of Quantum and Molecular Structures, Shanghai University, Shanghai, 200444, China}
\author{Junjie Liu}
\email{jj\_liu@shu.edu.cn}
\affiliation{Department of Physics, Institute for Quantum Science and Technology, Shanghai Key Laboratory of High Temperature Superconductors, International Center of Quantum and Molecular Structures, Shanghai University, Shanghai, 200444, China}

\begin{abstract}
The quantum Mpemba effect (QMPE), an anomalous relaxation phenomenon, has been demonstrated in both closed and open Hermitian quantum systems. While some studies have linked the QMPE to Liouvillian exceptional points--non-Hermitian features emerged at the Liouvillian level--in open Hermitian quantum systems, it remains largely unexplored whether the QMPE can occur in intrinsic non-Hermitian systems, where non-Hermiticity is inherent at the Hamiltonian level. Here, we demonstrate unequivocally the occurrence of QMPE in experimentally realizable parity-time-symmetric qubit systems immersed in a bosonic bath. Using established quantifiers for QMPE, we show numerically that the QMPE persists across parameter regimes both near and far from Hamiltonian and Liouvillian exceptional points, but disappears entirely when Hermitian Hamiltonian is restored. Interestingly, neither Hamiltonian nor Liouvillian exceptional points demarcate the boundaries of the QMPE regime. To complement numerical results, we develop an analytical description based on a long-time approximation of the relaxation dynamics of quantifiers. This approach allows us to decipher the number of intersections between two dynamical trajectories of quantifier starting from two initial conditions in the validity regime of the long-time approximation, thereby providing additional information that delineate the parameter regimes supporting the genuine QMPE. We further demonstrate the robustness of QMPE against increasing the number of qubits and dephasing effect. Our findings not only broaden the scope of the QMPE but also suggest its intricate interplay with non-Hermitian features beyond exceptional points.
\end{abstract}

\date{\today}
\maketitle

\section{Introduction}
The quantum Mpemba effect (QMPE) represents a fascinating extension of its classical counterpart--a counterintuitive thermal relaxation phenomenon where initially hotter liquids cool faster~\cite{EBMpemba_1969}. Recent theoretical studies have established the QMPE in diverse scenarios, including quantum thermal relaxation processes~\cite{Carollo.21.PRL,Kochsiek.22.PRA,Ivander.23.PRE,Moroder.24.PRL,Strachan.25.PRL,Van.25.PRL,Longhi.25.OL,Longhi.24.OL}, quantum battery~\cite{Medina.25.PRL}, quantum thermal machines~\cite{Lin.22.PRE,LiuD.24.PRA} and symmetry restoration~\cite{Ares.23.NC,LiuS.24.PRL,Turkeshi.25.PRL,Khor.24.Q,Rylands.24.JSM,Murciano.24.JSM,Ferro.24.JSM,Liu.24.A,Klobas.25.PRB,Yamashika.24.PRB} among others. Experimental verifications of the QMPE have also been achieved in several platforms~\cite{Zhang.25.NC,Joshi.24.PRL,Aharony.24.PRL,Xu.25.A}, providing conclusive evidence of the persistence of Mpemba effect in the quantum domain. Crucially, however, existing QMPE demonstrations remain largely restricted to Hermitian systems, whether in closed or open settings~\cite{Nava.19.PRB,Carollo.21.PRL,Kochsiek.22.PRA,Ivander.23.PRE,Moroder.24.PRL,Strachan.25.PRL,Van.25.PRL,Longhi.25.OL,Longhi.24.OL,Medina.25.PRL,Lin.22.PRE,LiuD.24.PRA,Ares.23.NC,LiuS.24.PRL,Turkeshi.25.PRL,Khor.24.Q,Rylands.24.JSM,Murciano.24.JSM,Ferro.24.JSM,Zhang.25.NC,Joshi.24.PRL,Aharony.24.PRL,Ares.25.NRP,Yu.25.APS,Dong.25.PRA,Qian.25.PRB,WangX.24.PRR,Nava.24.PRL,Furtado.25.AP,Kheirandish.24.A,Wang.24.A,Rylands.24.PRL,Xu.25.A,Liu.24.A,Ares.25.PRB,Klobas.25.PRB,Zatsarynna.25.PRB,Yamashika.24.PRB,Summer.25.A}. For a broader perspective, we refer readers to recent comprehensive reviews~\cite{Ares.25.NRP,Yu.25.APS}.

Uncovering distinct manifestations of the QMPE is critical to elucidating its underlying mechanisms. Non-Hermiticity, which profoundly modifies spectral properties, can host rich relaxation dynamics that are impossible in the Hermitian realm~\cite{Kawabata.17.PRL,Xiao.19.PRL,Wen.20.NPJQI,Fang.21.CP,Ding.22.PRA,Nori.22.PRR,Akram.23.SR,Ma.25.PRA,Guo.25.A}. In open Hermitian systems, non-Hermiticity emerges at the level of Liouvillian superoperators, giving rise to effective non-Hermitian systems~\cite{Giulio.25.E} or spectrum singularities known as Liouvillian exceptional points where both the eigenvalues and eigenmodes of Liouvillian superoperators coalesce~\cite{Prosen.10.JSM,Albert.14.PRA,Minganti.19.PRA,Hatano.19.MP,Minganti.18.PRA}. Recent works have revealed direct connections between the QMPE and Liouvillian exceptional points~\cite{Zhang.25.NC, Chatterjee.23.PRL,Chatterjee.24.PRA,Zhou.23.PRR}. In particular, a recent experiment demonstrated that Liouvillian exceptional points coincide precisely with the occurrence condition of the QMPE~\cite{Zhang.25.NC}, providing compelling evidence for the connection between the QMPE and non-Hermiticity. Motivated by this link, it is natural to ask whether the QMPE should generically appear in non-Hermitian systems. However, whether the QMPE can occur in intrinsic non-Hermitian systems, where non-Hermiticity is inherent at the Hamiltonian level and involves nontrivial non-Hermitian symmetries~\cite{Bender.1998.prl,bender.2005.cp,Bender.24.RMP,Ozdemir.19.NM}, remains largely unexplored. 
%-------------------------------------------------
\begin{figure}[t!]
 \centering
\includegraphics[width=1\columnwidth]{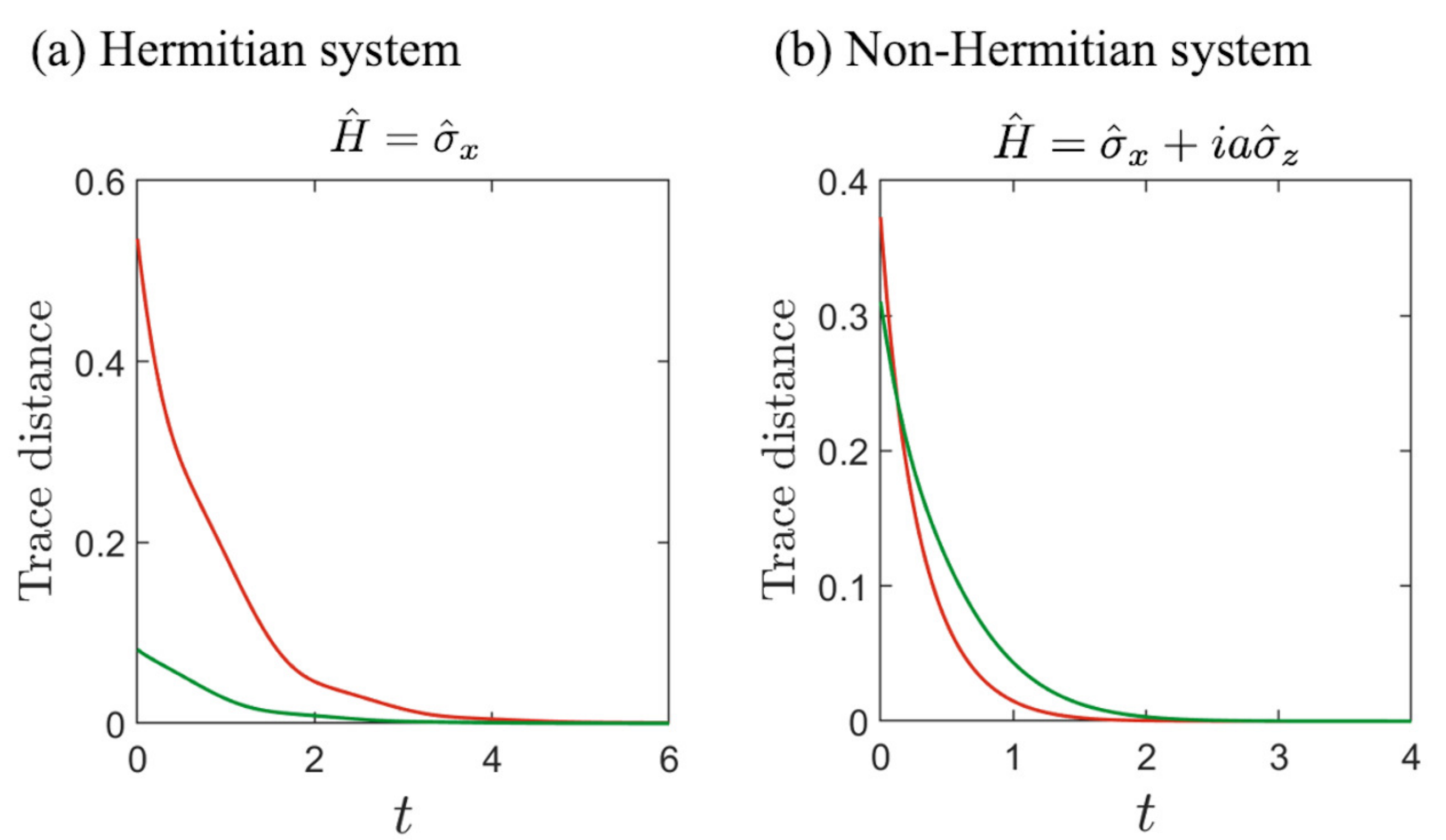} 
%\vspace*{-16mm}
\caption{The dynamical trajectories of the trace distance $D(t)$ [Eq. (\ref{trace_distance})] for (a) a Hermitian qubit system with Hamiltonian $\hat{H}=\hat{\sigma}_x$ showing no intersections, and (b) its PT-symmetric counterpart with Hamiltonian $\hat{H}=\hat{\sigma}_x+ia\hat{\sigma}_z$ ($a=1.2$) exhibiting a single intersection point indicating the QMPE. For both models: We obtain $D(t)$ by solving Eq. \eqref{rho(t)_Lindblad} under two different initial conditions $\hat{\rho}^{\rm I}(0)=\frac{1}{2}(\hat{\sigma}_z+\hat{\mathrm{I}})$ (red curves) and $\hat{\rho}^{\rm{II}}(0)=\frac{1}{2}\hat{\mathrm{I}}$ (green curves), where ``$\hat{\mathrm{I}}$" is the identity matrix. $\gamma_1=0.6$ and $\gamma_2=1$.}
\protect\label{fig:sketch}
\end{figure}
%-------------------------------------------------
The absence of this critical evidence has defied the ubiquity of the QMPE across Hermitian and non-Hermitian regimes--even amid strong indications~\cite{Zhang.25.NC, Chatterjee.23.PRL,Chatterjee.24.PRA,Zhou.23.PRR,Giulio.25.E,Guo.25.A}.

In this work, we investigate the QMPE in a parity-time-symmetric non-Hermitian qubit system coupled to a bosonic bath. Using established dynamical quantifiers~\cite{Ares.25.NRP}, we identify clear signatures of the QMPE, which vanishes completely as the non-Hermitian Hamiltonian is tuned to its Hermitian limit (see Fig. \ref{fig:sketch} for an illustration). In contrast to some prior studies on QMPE in open Hermitian systems, which highlighted a close connection between the QMPE and Liouvillian exceptional points~\cite{Zhang.25.NC,Chatterjee.24.PRA,Zhou.23.PRR}, we find that the QMPE regime in our system is not bounded by either Hamiltonian or Liouvillian exceptional points. This suggests that the relationship between the QMPE and exceptional points may be model-dependent rather than a general feature.

Interestingly, we show that the slowest relaxation mode in PT-symmetric qubit systems can be completely suppressed for a wide range of initial conditions, without requiring sophisticated initial state engineering~\cite{Carollo.21.PRL}. This suppression simplifies the analysis of relaxation dynamics and allows us to understand the occurrence of QMPE from a cooperative effect of residual relaxation modes~\cite{Chatterjee.23.PRL}. We exemplify this feature by deriving analytical expressions that determine the number of intersections between two dynamical trajectories of quantifiers starting from two initial conditions based on a long-time approximation of the relaxation dynamics. With numerical results and analytical expressions together, we can thus identify parameter regimes supporting genuine QMPE that only allows for an odd number of intersections. Furthermore, we confirm the robustness of QMPE against an increasing number of qubit and under a pure dephasing effect. We thus generalize the context of QMPE from previous Hermitian and effective non-Hermitian descriptions to intrinsic non-Hermitian systems, revealing distinct features that highlight the intricate interplay between non-Hermitian features beyond exceptional points and Mpemba phenomenon. Given recent realizations of PT-symmetric qubit systems~\cite{ChenW.21.PRL,Fang.21.CP,WangW.21.PRA}, our findings are testable within current experimental capabilities.

The structure of the paper is as follows. In Sec. \ref{sec:1}, we introduce the PT-symmetric non-Hermitian qubit model and the associated Lindblad-type quantum master equation that governs the evolution of a normalized density matrix. We also demonstrate that, under a relatively broad class of initial-state conditions, the slowest relaxation channel can be naturally and completely suppressed--without the need for any external control or initial state engineering.
In Sec. \ref{sec:2}, we examine the occurrence of the QMPE using several dynamical quantifiers, including the trace distance, Frobenius distance, and quantum relative entropy. We present numerical results for the number of intersections between different dynamical trajectories of the same quantifier starting from different initial conditions. We also obtain analytical boundaries for the parameter regimes exhibiting intersections based on a long-time approximation of dynamics. Together, we can identify parameter regimes supporting genuine QMPE. In Sec. \ref{sec:3}, we further investigate the robustness of the QMPE against dephasing effect and extend our analysis to PT-symmetric multiqubit systems. Finally, Sec. \ref{sec:4} summarizes our main results with some final remarks.

\section{Non-Hermitian qubit model}\label{sec:1}
To demonstrate the occurrence of the QMPE in intrinsic non-Hermitian systems, we consider an experimentally realizable PT-symmetric qubit system immersed in a bosonic thermal bath. The system Hamiltonian reads~\cite{ChenW.21.PRL,Fang.21.CP,WangW.21.PRA} 
\begin{equation}\label{eq:Hs}
    \hat H = \hat\sigma_x+ia\hat\sigma_z.
\end{equation}
Here, the operators $\hat{\sigma}_x$ and $\hat{\sigma}_z$ are Pauli matrices, $a$ governs the degree of non-Hermiticity. It can be readily checked that the above Hamiltonian is PT-symmetric, satisfying $[\hat{H},\hat{\mathcal{P}}\hat{\mathcal{T}}]=0$, with $\hat{\mathcal{P}}=\hat\sigma_x$ and $\hat{\mathcal{T}}$ denoting the parity and time-reversal operators, respectively. We note that $a=1$ marks a Hamiltonian exceptional point, separating a PT-unbroken ($0<a<1$) and a PT-broken ($a>1$) phases. By taking the Hermitian limit of $a=0$, we can get the Hermitian counterpart of $\hat H$. 

We consider normalized density matrix to describe the time evolution of non-Hermitian systems~\cite{Brody.12.PRL}. Under weak system-bath couplings, the evolution of the normalized density matrix $\hat{\rho}(t)$ of the open PT-symmetric qubit system can be governed by a Lindblad-type quantum master equation~\cite{Ma.25.PRA,Yuan.20.PRL,Xu.24.PRL} 
\bea
    \frac{d}{dt}\hat{\rho}(t) &=& -i\left[\hat{H}\hat{\rho}(t)-\hat{\rho}(t)\hat{H}^{\dagger}\right]+\sum_{k=1,2}\mathcal{D}_{k}[\hat{\rho}(t)]\nonumber\\
    &&-i\mathrm{Tr}[\hat{\rho}(t)(\hat{H}^{\dagger}-\hat{H})]\hat{\rho}(t).\label{rho(t)_Lindblad}
\eea
Here, the dissipator $\mathcal{D}_{k}[\hat{\rho}(t)]\equiv \gamma_{k}\left[\hat{L}_{k}\hat{\rho}(t)\hat{L}_{k}^{\dagger}-\frac{1}{2}\{\hat{L}_{k}^{\dagger}\hat{L}_{k},\hat{\rho}(t)\}\right]$ with $\gamma_k$ the decaying rate and $\hat{L}_k$ the Lindblad jump operator of dissipation channel $k$. $\{\hat{A}, \hat{B}\}$ denotes an anti-commutator between two operator $\hat{A}, \hat{B}$. The effect of a bosonic thermal bath is captured through two jump operators $\hat{L}_1=\hat{\sigma}_+$ and $\hat{L}_2=\hat{\sigma}_-$, where $\hat{\sigma}_{\pm}$ are spin ladder operators. The last term on the right-hand-side of Eq. \eqref{rho(t)_Lindblad} represents a correction term that ensures the conservation of probability during the evolution~\cite{Brody.12.PRL} and vanishes for Hermitian Hamiltonian. For derivation details that lead to Eq. (\ref{rho(t)_Lindblad}), we refer readers to Ref.~\cite{Ma.25.PRA}. 

In the absence of dissipator, Eq. (\ref{rho(t)_Lindblad}) describes the evolution of normalized density matrix of a closed non-Hermitian system, 
\begin{equation}
\hat{\rho}(t)=\frac{e^{-i\hat{H}t}\hat{\rho}(0)e^{i\hat{H}^\dagger t}}{\mathrm{Tr}[e^{-i\hat{H}t}\hat{\rho}(0)e^{i\hat{H}^\dagger t}]}.
\label{closed_H}
\end{equation}
Such a closed evolution has been realized experimentally~\cite{Fang.22.PRR,Xiao.19.PRL}. We also note that by tuning $a=0$, Eq. \eqref{rho(t)_Lindblad} reduces to the standard quantum Lindblad master equation describing a Hermitian qubit coupled to a thermal bath~\cite{Breuer.02.NULL}, with $\gamma_1/\gamma_2<1$ constrained by the detailed balance condition. To enable a physical Hermitian limit of Eq. \eqref{rho(t)_Lindblad}, we maintain the ratio $\gamma_1/\gamma_2<1$ for non-Hermitian qubit systems as well.

%-------------------------------------------------
\begin{figure}[t!] 
 \centering
\includegraphics[width=1\columnwidth]{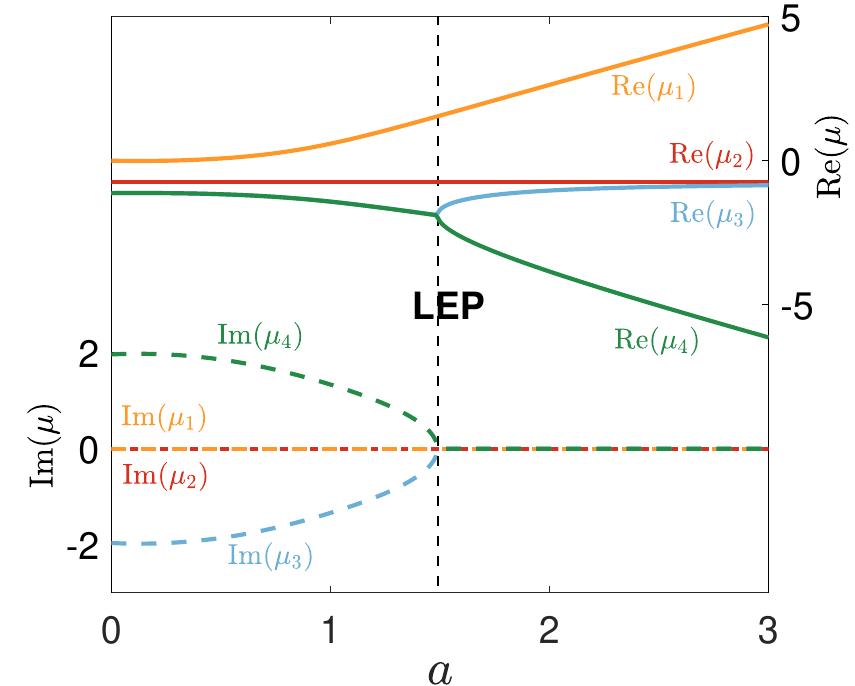} 
%\vspace*{-16mm}
\caption{Real (solid curves) and imaginary (dashed curves) parts of the eigenvalues $\{\mu_j\}$ of $\mathcal{L}_0$ for $\gamma_1=0.5$. The orange, red, blue, and green curves correspond to the eigenvalue branches $\mu_1$, $\mu_2$, $\mu_3$, and $\mu_4$, respectively. The eigenvalues remain non-degenerate throughout the parameter range except at the Liouvillian exceptional point (LEP, vertical dashed line), where two eigenvalues coalesce.}
\protect\label{fig:spectrum}
\end{figure}
%-------------------------------------------------
The formal solution of Eq. \eqref{rho(t)_Lindblad} can be expressed as $\hat{\rho}(t)=e^{\mathcal{L}_0t}\hat{\rho}(0)/\mathrm{Tr}\left[e^{\mathcal{L}_0t}\hat{\rho}(0)\right]$~\cite{Ma.25.PRA,Campaioli.24.PRXQ}, with the Liouvillian superoperator $\mathcal{L}_0$ reading $\mathcal{L}_0\hat{\rho}(t)\equiv-i\left[\hat{H}\hat{\rho}(t)-\hat{\rho}(t)\hat{H}^\dagger\right]+\sum_{k=1,2}\mathcal{D}_{k}[\hat{\rho}(t)]$. The denominator of the formal solution represents a normalization factor which originates from the correction term in Eq. \eqref{rho(t)_Lindblad}. By recasting $\mathcal{L}_0$ into a matrix form through vectorizing the density matrix, we can introduce the eigenspectrum of $\mathcal{L}_0$ through the relations $\mathcal{L}_0\hat{\rho}_j=\mu_j \hat{\rho}_j$ and $\mathcal{L}_0^{\dagger} \hat{\omega}_j=\mu_j^{\ast} \hat{\omega}_j$, where $\hat{\rho}_j$ and $\hat{\omega}_j$ correspond to the $j$th right and left {\it normalized} eigenvectors of $\mathcal{L}_0$ with eigenvalue $\mu_j$ and its complex conjugate $\mu_j^{\ast}$, respectively. We arrange them in descending order with respect to their real parts such that $\mathrm{Re}(\mu_1)\geqslant\mathrm{Re}(\mu_2)\geqslant\cdots$. We remark that the eigenspectrum of $\mathcal{L}_0$ exhibits a Liouvillian exceptional point (see a sketch in Fig. \ref{fig:spectrum}).

%-------------------------------------------------
%\begin{figure*}[tbh!]
 %\centering
%\includegraphics[width=2\columnwidth]{Spectrum.eps} 
%\vspace*{-16mm}
%\caption{(a) Real (solid curves) and imaginary (dashed curves) parts of the eigenvalues $\{\mu_j\}$ of $\mathcal{L}_0$ for $\gamma_1=0.5$. The vertical dashed line indicates the Liouvillian exceptional point (LEP). (b) Contour plot of the number of intersection points (non-negative integers) between two dynamical trajectories for $D^{\mathrm{I,II}}(t)$, calculated from Eq. (\ref{eq:rho_t}) under the initial conditions $\hat{\rho}^{\rm I}(0)=(\hat{\sigma}_z+\hat{\rm{I}})/2$ and $\hat{\rho}^{\rm{II}}(0)=\hat{\rm{I}}/2$, where $\hat{\rm{I}}$ is the identity matrix. Red dashed lines: analytical boundaries from Eq. (\ref{X=1}). Black dashed line: LEP. The orange star ($a=1,\gamma_1=0.4$) and red star ($a=1.3,\gamma_1=0.4$) mark parameters used in (c) and (d), respectively. (c) and (d) Detailed dynamics of $D^{\rm{I,II}}(t)$ for the parameters marked in (b). Insets: Difference $\Delta D(t)=D^{\rm I}(t)-D^{\rm{II}}(t)$ calculated from Eq. (\ref{eq:rho_t}) (solid line) and Eq. (\ref{rho_M}) (dash-dotted line); red circles indicate exact intersection points. The label $T=2.3$ in the inset of (c) denotes the theoretical period $\pi/|\mathrm{Im}(\mu_{3,4})|$. For all panels, $\gamma_2=1$.}
%\protect\label{fig:spectrum}
%\end{figure*}
%-------------------------------------------------

In terms of normalized eigenvectors and eigenvalues of $\mathcal{L}_0$, a time-dependent system density matrix $\hat{\rho}(t)$ can be decomposed as
\begin{equation}\label{eq:rho_t}
    \hat{\rho}(t)~=~\frac{\sum_je^{\mu_jt}C_j\hat{\rho}_j}{\mathrm{Tr}\left(\sum_je^{\mu_jt}C_j\hat{\rho}_j\right)}.
\end{equation}
Here, $C_j\equiv\mathrm{Tr}[\omega_j^\dagger\hat{\rho}(0)]$ marks the overlap between the initial system density matrix $\hat{\rho}(0)$ and $j$th relaxation mode, $\{\hat{\rho}_j\}$ appeared in the above decomposition are transformed into matrix forms. As $t\to\infty$, the normalized eigenvector $\hat{\rho}_1$ survives and represents the steady state $\hat{\rho}_{\rm{ss}}$ of the PT-symmetric qubit system immersed in a bosonic bath. We note that a conventional realization of the QMPE relies on suppressing the slowest relaxation mode through careful initial-state engineering with state-specific unitary transformations~\cite{Zhang.25.NC,Carollo.21.PRL, Moroder.24.PRL,Kochsiek.22.PRA}. Remarkably, in PT-symmetric qubit systems, such a suppression occurs naturally for a relatively wide range of initial states: For any incoherent (i.e., diagonal) initial states or coherent initial ones with purely imaginary off-diagonal elements, we have
\begin{equation}\label{C2}
    C_2=0,
\end{equation}
implying a complete suppression of the slowest relaxation mode without implementing sophisticated unitary transformations. For clarity, we relegate proof details to Appendix~\ref{a:1}. In light of conventional realization of the QMPE~\cite{Zhang.25.NC,Carollo.21.PRL, Moroder.24.PRL,Kochsiek.22.PRA}, one would expect the occurrence of the QMPE in such non-Hermitian systems by contrasting the dynamical trajectory starting from an initial state within the aforementioned set satisfying Eq. (\ref{C2}) with that from a coherent initial state outside the set--specifically, one with complex off-diagonal elements leading to $C_2\neq 0$. However, here we explore whether the QMPE can still occur when both initial conditions satisfy $C_2=0$. If confirmed, such a manifestation would stem from a cooperative effect among the remaining relaxation modes, implying a more intricate underlying mechanism. To systematically examine this scenario, we restrict our subsequent analysis to initial states fulfilling Eq. (\ref{C2}).

\section{Identifying quantum Mpemba effect}\label{sec:2}
To establish an unambiguous signature of the QMPE, we contrast the time evolution of the PT-symmetric qubit system initialized in two distinct states $\hat{\rho}^{\rm{I},\rm{II}}(0)$ with different initial distances from the steady state $\hat{\rho}_{\rm{ss}}$~\cite{Nava.24.PRL}. An intersection between two dynamical trajectories occurs when any valid dynamical quantifier $\mathcal{O}(t)$ fulfills
\begin{equation}\label{MPE_condition}
    \mathcal{O}^{\mathrm{I}}(0)\neq \mathcal{O}^{\mathrm{II}}(0) \quad\mathrm{and}\quad \mathcal{O}^{\mathrm{I}}(\tau)=\mathcal{O}^{\mathrm{II}}(\tau).
\end{equation}
Here, the superscripts reflect the chosen initial states, $\tau$ marks an intersection point between two dynamical trajectories of $\mathcal{O}^{\mathrm{I},\rm{II}}(t)$. Typically, there is a sign change of $\mathcal{O}^{\mathrm{I}}(t)- \mathcal{O}^{\mathrm{II}}(t)$ before and after the intersection time point $\tau$. We note that multiple intersections may occur between dynamical trajectories~\cite{Ares.23.NC,Chatterjee.23.PRL,Chatterjee.24.PRA,WangX.24.PRR,Chalas.24.JSP}. In such cases, a genuine QMPE is identified only when the number of intersections is odd. In the following analysis, we explicitly employ the trace distance~\cite{Nava.24.PRL,Ares.25.NRP,Nielsen.00.NULL}, Frobenius distance~\cite{Carollo.21.PRL}, and quantum relative entropy~\cite{Chatterjee.23.PRL,Moroder.24.PRL} as dynamical quantifiers.

%-------------------------------------------------
\begin{figure}[t!] 
 \centering
\includegraphics[width=1\columnwidth]{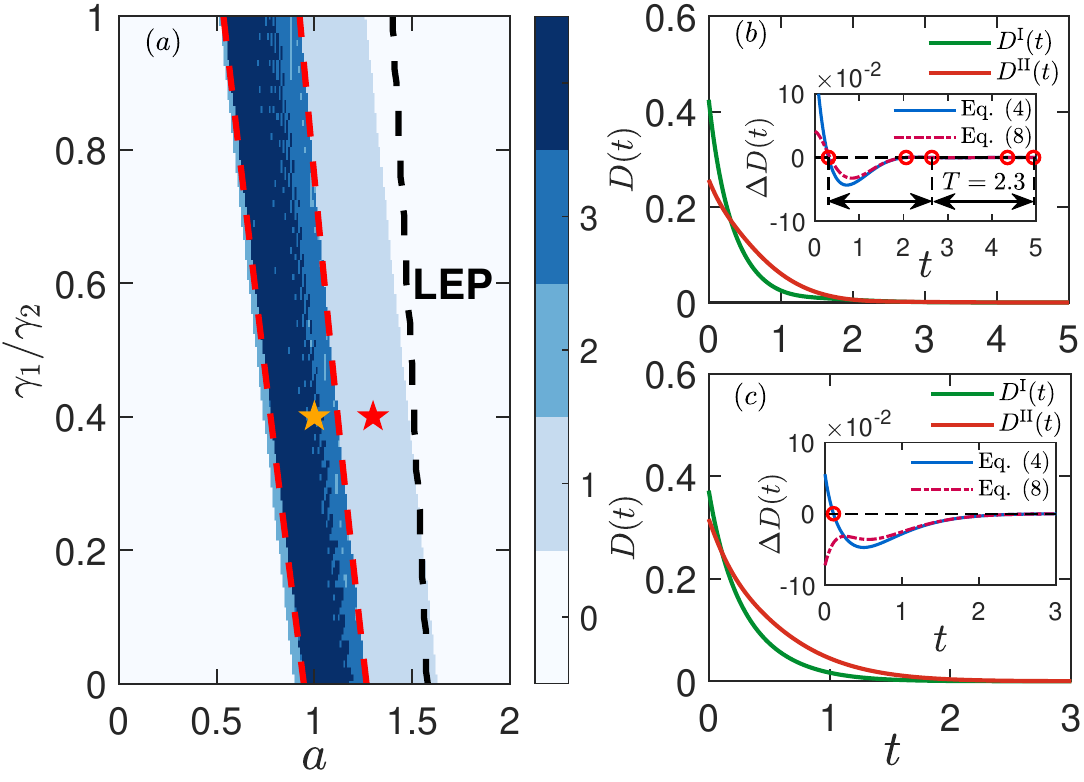} 
%\vspace*{-16mm}
\caption{(a) Contour plot of the number of intersection points (non-negative integers) counted up to $t=20$ between two dynamical trajectories for $D^{\mathrm{I,II}}(t)$, calculated from Eq. (\ref{eq:rho_t}) under the initial conditions $\hat{\rho}^{\rm I}(0)=(\hat{\sigma}_z+\hat{\rm{I}})/2$ and $\hat{\rho}^{\rm{II}}(0)=\hat{\rm{I}}/2$, where $\hat{\rm{I}}$ is the identity matrix. Red dashed lines: analytical boundaries from Eq. (\ref{X=1}) (see more details in Fig. \ref{fig:x_trace_left} and main text). Black dashed line: LEP. The orange star ($a=1,\gamma_1=0.4$) and red star ($a=1.3,\gamma_1=0.4$) mark parameters used in (b) and (c), respectively. (b) and (c) Detailed dynamics of $D^{\rm{I,II}}(t)$ for the parameters marked in (a). Insets: Difference $\Delta D(t)=D^{\rm I}(t)-D^{\rm{II}}(t)$ calculated from Eq. (\ref{eq:rho_t}) (solid line) and Eq. (\ref{rho_M}) (dash-dotted line); red circles indicate exact intersection points. The label $T=2.3$ in the inset of (c) denotes the theoretical period $\pi/|\mathrm{Im}(\mu_{3,4})|$. For all panels, $\gamma_2=1$.}
\protect\label{fig:long_trace_left}
\end{figure}
%-------------------------------------------------
\subsection{Trace distance}
We employ the trace distance~\cite{Nava.24.PRL,Ares.25.NRP,Nielsen.00.NULL}
\begin{equation}\label{trace_distance}
D(t)=\frac{1}{2}\mathrm{Tr}|\hat{\rho}(t)-\hat{\rho}_{\mathrm{ss}}|
\end{equation}
as our first dynamical quantifier for the QMPE identification, which measures the distance between the time-evolving state $\hat{\rho}(t)$ and the final steady state $\hat{\rho}_{\rm{ss}}$. Here, $|\hat{A}|\equiv\sqrt{\hat{A}^{\dagger}\hat{A}}$. This dynamical measure has been used experimentally to probe the QMPE~\cite{Zhang.25.NC,Aharony.24.PRL}. To demonstrate the occurrence of the QMPE in PT-symmetric qubit systems, we numerically simulate and contrast two dynamical trajectories of the trace distance $D^{\rm{I,II}}(t)$ by evolving Eq. (\ref{eq:rho_t}) under two different initial conditions $\hat{\rho}^{\rm{I},\rm{II}}(0)$. Our comprehensive simulations explore the parameter space spanned by $a$ and $\gamma_{1,2}$ subject to the constraint $\gamma_1/\gamma_2<1$, and track the emergence of intersection points between the two dynamical trajectories of $D^{\rm I}(t)$ and $D^{\rm{II}}(t)$ as captured by the conditions in Eq. (\ref{MPE_condition}).

Fig. \ref{fig:long_trace_left} (a) presents a numerical contour plot of the number of intersection points counted up to $t=20$; Intersection points beyond $t=20$ are numerically challenging to identify as magnitudes of both $D^{\rm {I,II}}(t)$ are too small. The contour plot reveals well-separated parameter regimes exhibiting either single or multiple intersection points. The emergence of multiple intersections aligns with earlier findings in both closed and open Hermitian quantum systems~\cite{Ares.23.NC,Chatterjee.23.PRL,Chatterjee.24.PRA,WangX.24.PRR,Chalas.24.JSP}. To complement numerical results for intersection counting, we show in Fig. \ref{fig:long_trace_left} (b) and (c) detailed dynamical trajectories of $D^{\rm{I,II}}(t)$, computed from Eq. (\ref{eq:rho_t}) under two different initial conditions $\hat{\rho}^{\rm{I,II}}(0)$. As we adopt parameters marked by the orange and red stars in Fig. \ref{fig:long_trace_left} (a) for Fig. \ref{fig:long_trace_left} (b) and (c), the two dynamical trajectories depict multiple and single intersection points, respectively, as highlighted by the marked zero points (red circles) of $\Delta D(t)=D^{\rm{I}}(t)-D^{\rm{II}}(t)$ (blue curves) in the insets. Notably, the single intersection point in Fig. \ref{fig:long_trace_left} (c) occurs at relatively short times, whereas multiple intersection points in Fig. \ref{fig:long_trace_left} (b) persist to much longer time scales.

Since numerical counting becomes infeasible at long times, it is difficult to conclusively identify or rule out the occurrence of the QMPE in this system based solely on numerical results. We therefore resort to analytical treatments to reliably determine the parameter regimes in which the QMPE emerges, characterized by an odd number of intersections between dynamical trajectories. Noting the complete absence of the slowest relaxation mode under our initial conditions, we introduce a long-time approximation to the normalized density matrix given in Eq. (\ref{eq:rho_t}) (see details in Appendix~\ref{a:2})
\begin{equation}
\hat{\rho}(t)~\simeq~\hat{\rho}_{\rm{ss}}+\hat{M}(t)e^{-(\mu_{1}-\mu_{3})t},\label{rho_M}
\end{equation}
where $\hat{M}(t)=\frac{C_{3}}{C_{1}}(\hat{\rho}_{3}-\hat{\rho}_{1})+\frac{C_{4}}{C_{1}}e^{-(\mu_{3}-\mu_{4})t}\left(\hat{\rho}_{4}-\hat{\rho}_{1}\right)$ reflects the interplay between remaining relaxation modes. One can check that $\hat{M}(t)e^{-(\mu_{1}-\mu_{3})t}$ as a whole is Hermitian by using spectrum properties presented in Appendix~\ref{a:1}. We emphasize that Eq. (\ref{rho_M}), in conjunction with numerical results, enables a systematic delineation of the parameter regimes supporting the QMPE.

To see this, we insert Eq. (\ref{rho_M}) into Eq. (\ref{trace_distance}) and obtain 
\begin{equation}\label{eq:DD_sim}
D(t)~\simeq~\frac{1}{2}\mathrm{Tr}|\hat{M}(t)e^{-(\mu_{1}-\mu_{3})t}|.
\end{equation}
We note that $\hat{M}(t)$ is a traceless matrix as $\{\hat{\rho}_i\}$ are normalized eigenmatrices in our treatment such that $\mathrm{Tr}[\hat{\rho}_i]=1$, implying that two eigenvalues of $\hat{M}(t)$ have the same magnitude. Hence, the trace norm $\mathrm{Tr}|\hat{M}(t)e^{-(\mu_{1}-\mu_{3})t}|=2|\lambda(t)||e^{-(\mu_{1}-\mu_{3})t}|$ with $\lambda(t)$ being one of two eigenvalues of the matrix $\hat{M}(t)$. We now employ Eq. (\ref{eq:DD_sim}) to determine whether--and under which conditions--the two dynamical trajectories of of $D(t)$ intersect. Assuming an intersection at time $\tau$ exists, we combine Eq. (\ref{MPE_condition}) and Eq. (\ref{eq:DD_sim}) to obtain $|\lambda^{\mathrm{I}}(\tau)|=|\lambda^{\mathrm{II}}(\tau)|$, which is equivalent to 
\begin{equation}\label{M_trace}
\mathrm{Tr}\left([\hat{M}^{\mathrm{I}}(\tau)]^2\right)~=~\mathrm{Tr}\left([\hat{M}^{\mathrm{II}}(\tau)]^2\right)
\end{equation}
for a traceless matrix. Since $\mathrm{Tr}\left([\hat{M}^{i}(\tau)]^2\right)=(R_3^{i})^{2}T_3+2R_3^{i}R_4^{i}PX(\tau)+(R_4^{i})^{2}T_4X^{2}(\tau)$ ($i=\mathrm{I},\mathrm{II}$) with $T_3=\mathrm{Tr}[(\hat{\rho}_3-\hat{\rho}_1)^2]$, $T_4=\mathrm{Tr}[(\hat{\rho}_4-\hat{\rho}_1)^2]$,$P=\mathrm{Tr}[(\hat{\rho}_3-\hat{\rho}_1)(\hat{\rho}_4-\hat{\rho}_1)]$, $R_3^i=C_3^i/C_1^i$, $R_4^i=C_4^i/C_1^i$ and $X(\tau)=e^{-(\mu_{3}-\mu_{4})\tau}$, Eq. (\ref{M_trace}) reduces to a quadratic equation for $X(\tau)$ with two solutions
\begin{equation}
[X(\tau)]_{\pm}=\frac{-2P(R_3^{\mathrm{I}}R_4^{\mathrm{I}}-R_3^{\mathrm{II}}R_4^{\mathrm{II}})\pm\sqrt{\Delta}}{2T_4 \left[ (R_4^{\mathrm{I}})^2-(R_4^{\mathrm{II}})^2 \right]},\label{X_solution}
\end{equation}
where $\Delta\equiv4P^2(R_3^{\mathrm{I}}R_4^{\mathrm{I}}-R_3^{\mathrm{II}}R_4^{\mathrm{II}})^2-4T_3T_4[(R_3^{\mathrm{I}})^{2}-(R_3^{\mathrm{II}})^{2}][(R_4^{\mathrm{I}})^{2}-(R_4^{\mathrm{II}})^{2}]$. 
We emphasize that the two solutions for $X(\tau)$, which are fully determined by the spectrum of $\mathcal{L}_0$ and initial conditions, encode the possible intersection time $\tau$. However, we note that the solutions in Eq. (\ref{X_solution}) cannot always guarantee a physical $\tau$ which must be both real and positive, since $X(\tau)$ also depends on $\mu_{3,4}$--eigenvalues of the Liouvillian superoperator $\mathcal{L}_0$.

%-------------------------------------------------
\begin{figure}[t!] 
 \centering
\includegraphics[width=1\columnwidth]{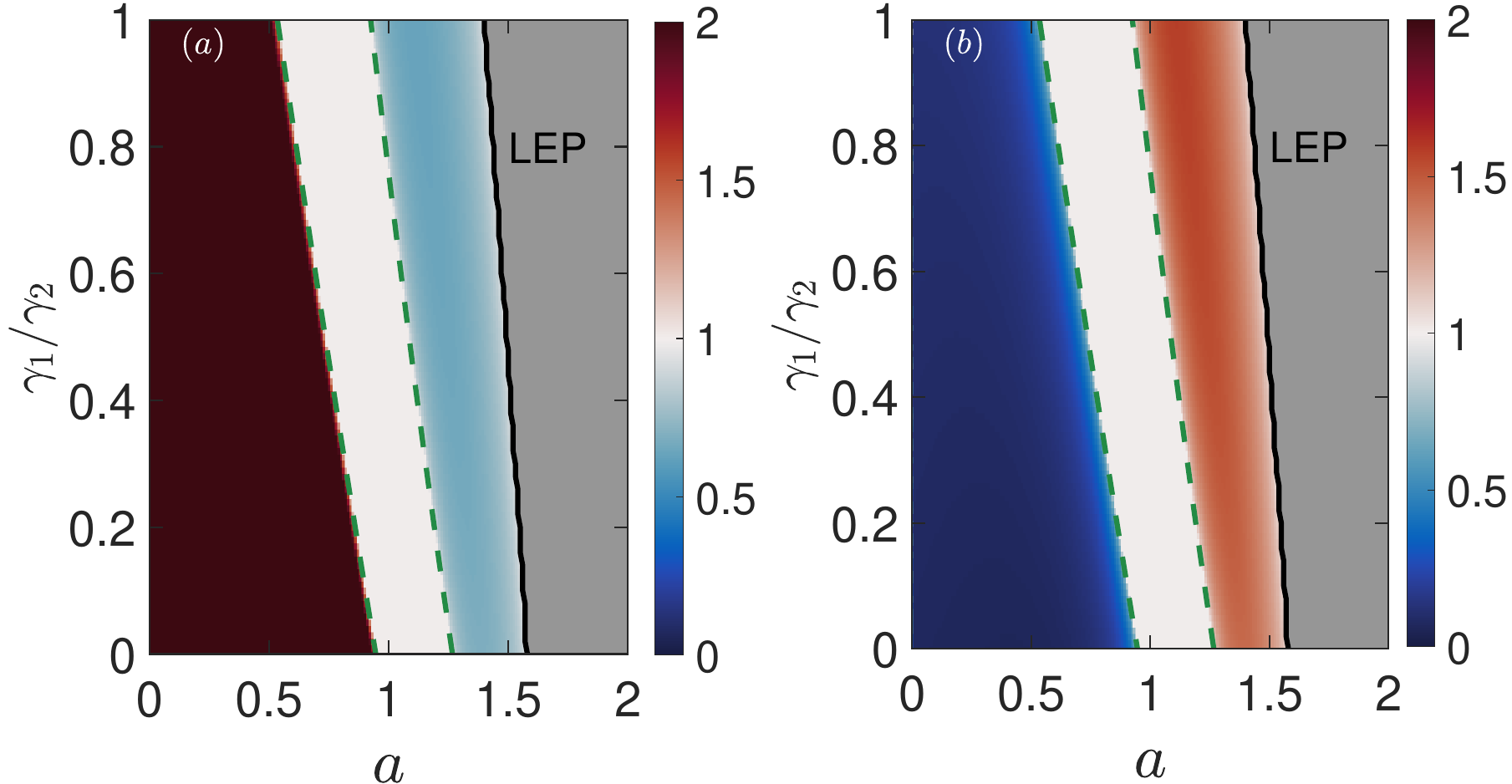} 
%\vspace*{-16mm}
\caption{Contour plot of magnitudes of theoretical solutions: (a) $|[X(\tau)]_{+}|$ and (b) $|[X(\tau)]_{-}|$ given in Eq. (\ref{X_solution}) in the regime to the left of Liouvillian exceptional point (LEP, black solid lines). Green dashed lines are theoretical boundaries for the white shaded regime that validates the constraint in Eq. (\ref{X=1}) and are used as red dashed lines in Fig. \ref{fig:long_trace_left} (a). Gray shaded regime to the right of the LEP is not considered. $\gamma_2=1$ for all plots. We adopt the same initial conditions as Fig. \ref{fig:long_trace_left}.}
\protect\label{fig:x_trace_left}
\end{figure}
%-------------------------------------------------

To proceed, we should further take into account the spectrum properties of $\mu_{3,4}$. 
From Fig. \ref{fig:long_trace_left} (a), we observe that intersections occur to the left of the Liouvillian exceptional point, corresponding to the parameter range where $\mathcal{L}_0$ exhibits complex eigenvalues (see Fig. \ref{fig:spectrum}), with particularly, $\mu_{3,4}$ forming a complex conjugate pair. Recalling that $X(\tau)=e^{-(\mu_{3}-\mu_{4})\tau}$, the requirement of a real-valued $\tau$ implies that $[X(\tau)]_{\pm}$ must be periodic functions of $\tau$, with a period $T=\pi/|\mathrm{Im}[\mu_{3,4}]|$ determined by the imaginary part of $\mu_{3,4}$. This periodicity further indicates that multiple intersection points occur periodically with period $T$ to the left of the Liouvillian exceptional point. This theoretical predicted periodicity is verified by our finite-time numerical results as evident from the comparison showed in the inset of Fig. \ref{fig:long_trace_left} (b). Hence, to ensure that Eq. (\ref{X_solution}) yields a physically admissible $\tau$ in the regime to the left of the Liouvillian exceptional point, we must further enforce the following additional constraint
\begin{equation}
    \left|[X(\tau)]_{\pm}\right|=1.\label{X=1}
\end{equation}
This condition follows naturally from the magnitude of a periodic function $X(\tau)=e^{-(\mu_{3}-\mu_{4})\tau}$. When Eq. (\ref{X=1}) does not hold for the two solutions, a real $\tau$ is no longer guaranteed.

We expect that Eqs. (\ref{X_solution}) and (\ref{X=1}) jointly determine the parameter regime exhibiting multiple intersection points for $a<a_{\rm{LEP}}$. To validate this theoretical analysis against numerical results, we systematically vary parameters $a$ and $\gamma_{1,2}$, compute the values of $|[X(\tau)]_{\pm}|$ using the expressions in Eq. (\ref{X_solution}) with spectrum of $\mathcal{L}_0$ and initial conditions as input, and assess whether they satisfy the constraint specified in Eq. (\ref{X=1}). In Fig. \ref{fig:x_trace_left} (a) and (b), we depict contour plots of the magnitudes of theoretical solutions $|[X(\tau)]_{+}|$ and $|[X(\tau)]_{-}|$ given in Eq. (\ref{X_solution}), respectively. From the figure, we first observe that the two conditions $|[X(\tau)]_{\pm}|=1$ [Eq. (\ref{X=1})] are satisfied within the same white shaded region, consistent with Eq. (\ref{X_solution})~\footnote{With the solutions in Eq. (\ref{X_solution}), we find $X_+ X_-=\frac{T_3\left[(R_3^\mathrm{I})^2-(R_3^\mathrm{II})^2\right]}{T_4\left[(R_4^\mathrm{I})^2-(R_4^\mathrm{II})^2\right]}=\frac{T_3\left[(R_3^\mathrm{I})^2-(R_3^\mathrm{II})^2\right]}{T_3^{\ast}\left[(R_3^\mathrm{I})^2-(R_3^\mathrm{II})^2\right]^{\ast}}$ due to properties that $C_1$ is strictly real and $C_{3,4}$ form a complex conjugate pair irrespective of initial conditions (Appendix \ref{a:1}). Hence, we have $|X_+ X_-|=1$, implying that $|X_+|=1$ when $|X_-|=1$ and vice versa.}. Outside this white shaded region--whose boundaries are indicated by green dashed lines (corresponding to the red dashed lines in Fig. \ref{fig:long_trace_left})--we have $|[X(\tau)]_{\pm}|\neq 1$ in general, implying that $\tau$ is no longer real. We therefore identify the white shaded region as the physical regime supporting multiple intersections. This theoretical region coincides precisely with the numerically identified regime exhibiting multiple intersection points as can been seen from Fig. \ref{fig:long_trace_left} (a) and (b). However, as our numerical simulations are necessarily restricted to finite times and intersections at late stages become difficult to resolve, the number of intersections identified numerically remains finite. This stands in contrast to the ideal case of an infinite number of intersections predicted by the theoretical analysis. Nevertheless, it is safe to exclude this regime of multiple intersections as one that supports the QMPE.

By contrast, we note that the regime exhibiting a single intersection point at short times lies beyond our theoretical boundaries, as the long-time approximation in Eq. (\ref{eq:DD_sim}) cannot capture the short dynamics in this regime. This distinction is demonstrated explicitly in the inset of Fig. \ref{fig:long_trace_left} (c), where we compare the difference $\Delta D(t)=D^{\rm I}(t)-D^{\rm{II}}(t)$ with $D^{\rm{I,II}}(t)$ calculated using either the full form of the normalized density matrix [cf. Eq. (\ref{eq:rho_t})] or its long-time approximation [cf. Eq. (\ref{rho_M})]. However, our long-time approximation in Eq. (\ref{eq:DD_sim}) accurately captures the late-time dynamics of $\Delta D(t)$ in this regime, as illustrated in the inset of Fig. \ref{fig:long_trace_left} (c). Using Eq. (\ref{eq:DD_sim}), we confirm that no additional intersections occur at large times, indicating that the relaxation process features only a single intersection at short times. This regime can therefore be identified as supporting the genuine QMPE.

From Fig. \ref{fig:long_trace_left} (a), we further note that the parameter regime where the QMPE occurs is not bounded by exceptional points. Specifically, neither the Hamiltonian exceptional point ($a=1$) nor the Liouvillian exception point $a_{\rm {LEP}}$ (black dashed line in Fig. \ref{fig:long_trace_left} (a)) defines their boundaries. This suggests that there is no connection between exceptional points and the QMPE in PT-symmetric qubit system. This observation contrasts with findings in open Hermitian systems~\cite{Zhang.25.NC,Chatterjee.24.PRA,Zhou.23.PRR}. Moreover, we find from Fig. \ref{fig:long_trace_left} that the parameter regime exhibiting the QMPE does not extend to the Hermitian limit with $a=0$ and $\gamma_1/\gamma_2<1$, indicating that the observed QMPE represents a phenomenon of solely PT-symmetric qubit system with $a\neq 0$. The absence of QMPE in the Hermitian limit can be understood from Fig. \ref{fig:x_trace_left}, which shows that $|[X(\tau)]_{\pm}|\neq 1$ in that limit, corresponding to a non-physical intersection time $\tau$.

%-------------------------------------------------
\begin{figure}[t!]
 \centering
\includegraphics[width=1\columnwidth]{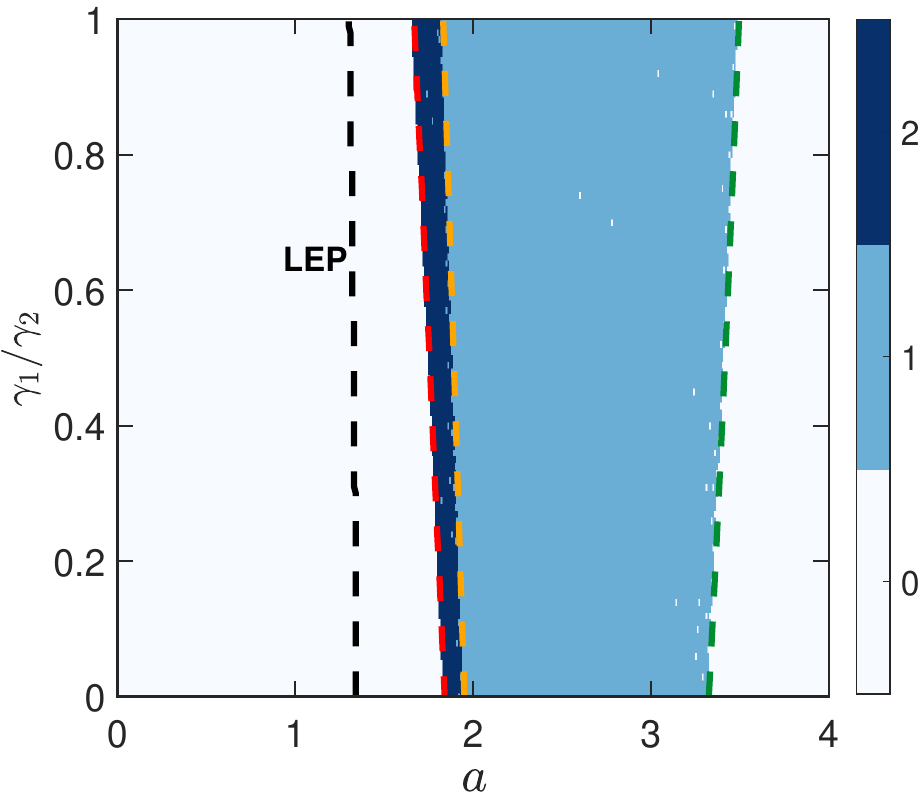} 
%\vspace*{-16mm}
\caption{Contour plot of the number of intersection points (non-negative integers) counted up to $t=20$ between two dynamical trajectories for $D^{\mathrm{I,II}}(t)$, calculated from Eq. (\ref{eq:rho_t}) under the initial conditions $\hat{\rho}^{\rm I}(0)=\frac{1}{2}\hat{\mathrm{I}}-0.3\hat{\sigma}_z-0.2\hat{\sigma}_y$ and $\hat{\rho}^{\rm{II}}(0)=\frac{1}{2}\hat{\mathrm{I}}$. The red, orange and green dashed lines are theoretical boundaries predicted by Eq. (\ref{X_01}) (see more details in Fig. \ref{fig:x_trace_right} and main text). The black dashed line denotes the position of LEP. $\gamma_2=0.5$.
}
\protect\label{fig:full_trace}
\end{figure}
%-------------------------------------------------
We remark that the QMPE can also occur in the regime to the right of the Liouvillian exceptional point. Fig. \ref{fig:full_trace} shows a representative contour plot of the number of intersection points counted up to $t=20$, revealing distinct parameter regimes with either one or two intersections. To determine whether additional intersections occur at long times, we again resort to solutions $[X(\tau)]_{\pm}$ from Eq. (\ref{X_solution}), which theoretically determine the positions and possible number of intersections. Similarly, we should identify constraints on $[X(\tau)]_{\pm}$ that ensure a physical intersection time $\tau$. From Fig. \ref{fig:spectrum}, we know that the Liouvillian's eigenvalues $\{\mu_j\}$ become purely real when $a>a_{\rm{LEP}}$, rendering $X(\tau)=e^{-(\mu_{3}-\mu_{4})\tau}$ a real function. To ensure $\tau>0$, we must impose the condition  
\begin{equation}\label{X_01}
    0<[X(\tau)]_{\pm}<1, 
\end{equation}
which guarantees two real and positive solutions for the intersection time, $\tau_{\pm}=-\ln([X(\tau)]_{\pm})/(\mu_3-\mu_4)$ (Recalling that we arrange eigenvalues in descending order). 

%-------------------------------------------------
\begin{figure}[b!] 
 \centering
\includegraphics[width=1\columnwidth]{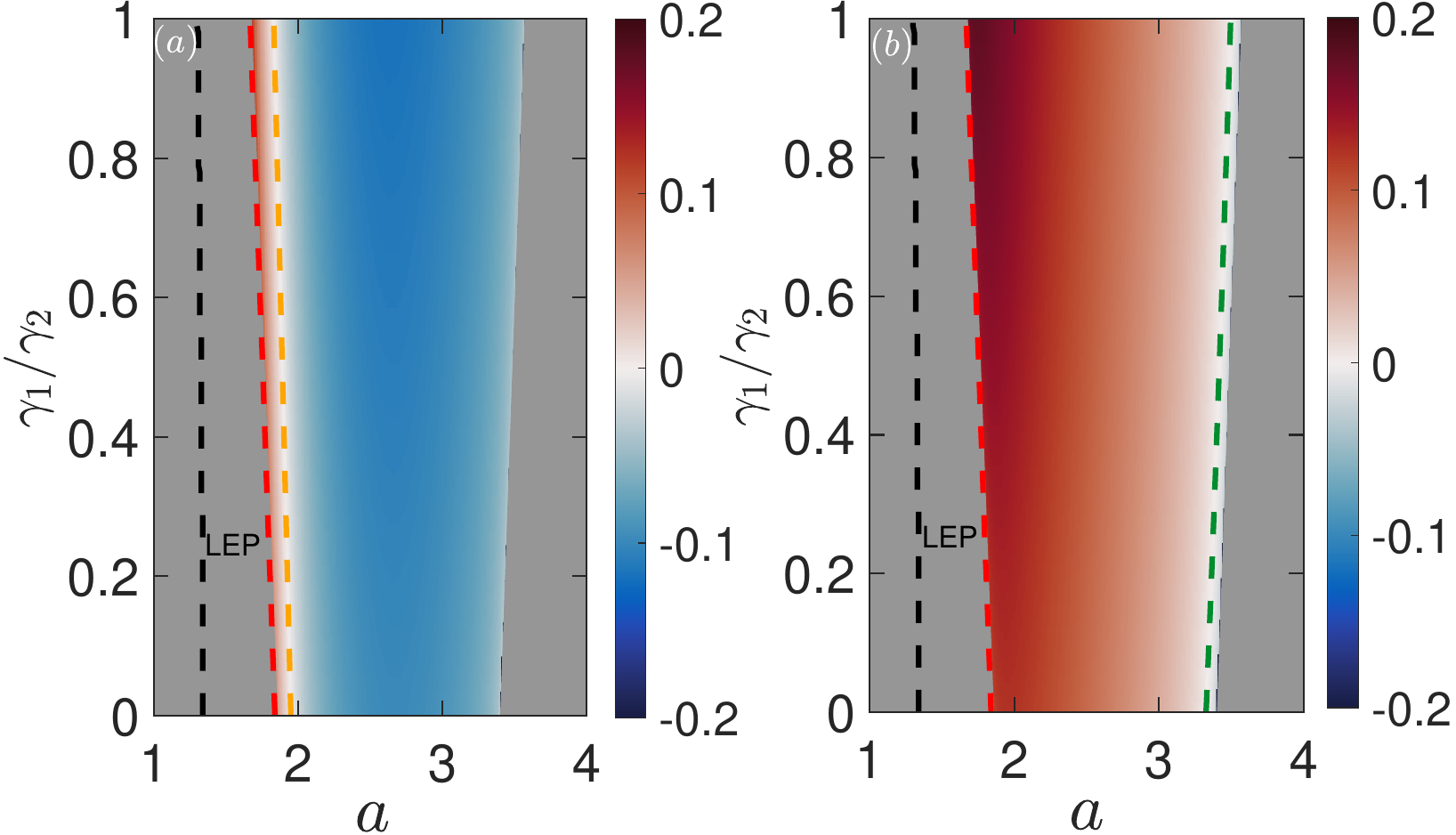} 
%\vspace*{-16mm}
\caption{Contour plot of magnitudes of theoretical solutions: (a) $[X(\tau)]_{+}$ and (b) $[X(\tau)]_{-}$ given in Eq. (\ref{X_solution}) in the regime to the right of Liouvillian exceptional point (LEP, black dashed lines). Red, orange and green dashed lines are boundaries for the shaded regimes that validate the constraint Eq. (\ref{X_01}) and are used in Fig. \ref{fig:full_trace}. Gray shaded regimes depict complex $[X(\tau)]_{\pm}$ and are thus not considered. $\gamma_2=1$ for all plots. We adopt the same initial conditions as Fig. \ref{fig:full_trace}.}
\protect\label{fig:x_trace_right}
\end{figure}
%-------------------------------------------------
We now determine the parameter regimes in which Eq. (\ref{X_01}) is satisfied. We still vary parameters $a$ and $\gamma_{1,2}$, compute the values of $[X(\tau)]_{\pm}$ using the expressions in Eq. (\ref{X_solution}) with spectrum of $\mathcal{L}_0$ and given initial conditions as input, and assess whether they satisfy the constraint specified in Eq. (\ref{X_01}). In Fig. \ref{fig:x_trace_right} (a) and (b), we present contour plots of the values of $[X(\tau)]_{+}$ and $[X(\tau)]_{-}$, respectively. From Fig. \ref{fig:x_trace_right} (a), we observe that the condition $0<[X(\tau)]_{+}<1$ is satisfied only within a narrow region bounded by red and orange dashed lines. In comparison, Fig. \ref{fig:x_trace_right} (b) shows that the condition $0<[X(\tau)]_{-}<1$ holds within a broader region delineated by red and green dashed lines. Notably, $0<[X(\tau)]_{-}<1$ is also satisfied in the region of $0<[X(\tau)]_{+}<1$. Outside these regimes satisfying Eq. (\ref{X_01}), we generally encounter two scenarios: (i) $[X(\tau)]_{\pm}$ remain real but take negative values (blue shaded region), or (ii) $[X(\tau)]_{\pm}$ become complex (gray shaded regions). In neither case is a real and non-negative $\tau$ ensured. The validity of these theoretical boundaries is confirmed in Fig. \ref{fig:full_trace} by comparing them with numerical identified regimes exhibiting intersections. Notably, the regime between the red and orange dashed lines in Fig. \ref{fig:full_trace} satisfies both conditions in Eq. (\ref{X_01}). The perfect agreement between the theoretical boundaries and the numerically identified regimes confirms that: (i) the transition from one to two intersection points observed numerically in Fig. \ref{fig:full_trace} as $a$ decreases is well captured, and (ii) no additional intersections occur at large times. Therefore, the region between the orange and green dashed lines in Fig. \ref{fig:full_trace}, which exhibits a single intersection, can be identified as supporting the genuine QMPE.

\subsection{Frobenius distance}
In addition to trace distance, the Frobenius distance offers an alternative and computationally convenient way to quantify the difference between a quantum state and its steady state. For a time-evolved density matrix $\hat{\rho}(t)$ and the steady state $\hat{\rho}_{\mathrm{ss}}$, the Frobenius distance is defined as~\cite{Carollo.21.PRL,Ares.25.NRP}
\begin{equation}
    D_{\mathrm{F}}(t)=\sqrt{\mathrm{Tr}\hat{A}^{\dagger}(t)\hat{A}(t)}.
\end{equation}
Here, we have defined $\hat{A}(t)\equiv\hat{\rho}(t)-\hat{\rho}_\mathrm{ss}=\hat{A}^{\dagger}(t)$. This measure captures the overall distance between two matrices in Hilbert–Schmidt space and is sensitive to both population and coherence differences.

%-------------------------------------------------
\begin{figure}[b!]  
 \centering
\includegraphics[width=1\columnwidth]{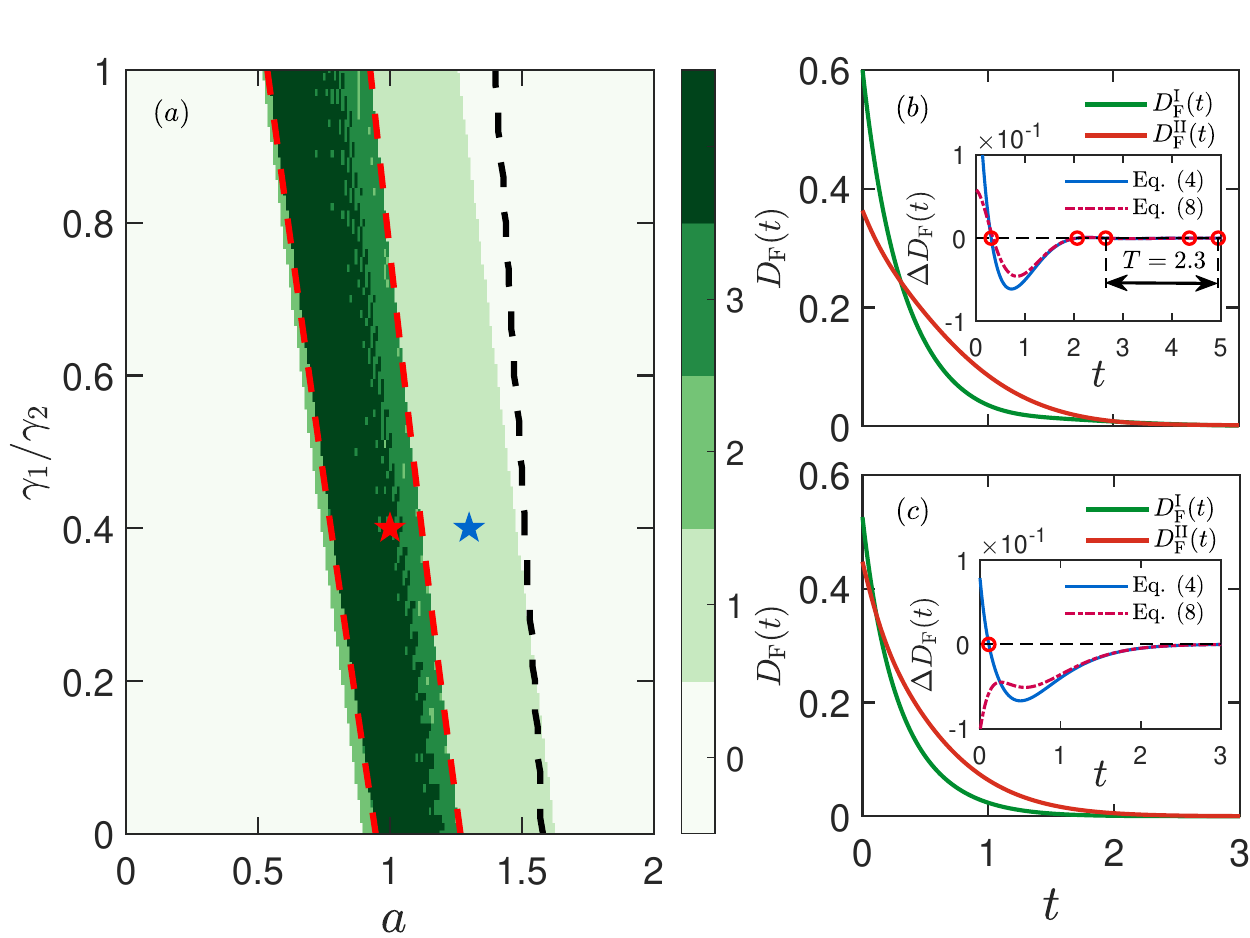} 
%\vspace*{-16mm}
\caption{(a) Contour plot of intersection point counts (non-negative integers) counted up to $t=20$ between two dynamical trajectories of $D_{\rm F}(t)$ calculated using the full form of $\hat{\rho}(t)$ under the given two initial conditions. The red dashed lines are analytical boundaries predicted by Eq. (\ref{X=1}) (see also Fig. \ref{fig:x_trace_left}). The black dashed line denotes the position of LEP. The red ($a=1,\gamma_1=0.4$) and blue ($a=1.3,\gamma_1=0.4$) stars mark the parameter sets used in (b) and (c) which depict dynamical trajectories of $D_{\rm F}^{\rm{I,II}}(t)$. Insets in (b) and (c): Insets: Difference $\Delta D_{\rm F}(t)=D_{\rm F}^{\rm I}(t)-D_{\rm F}^{\rm{II}}(t)$ calculated from Eq. (\ref{eq:rho_t}) (solid line) and Eq. (\ref{rho_M}) (dash-dotted line); red circles indicate exact intersection points. $T=2.3$ in inset of (b) denotes the theoretical prediction of period $\pi/|\mathrm{Im}[\mu_{3,4}]|$. $\gamma_2=1$ for all plots.}
\protect\label{fig:full_Fd}
\end{figure}
%-------------------------------------------------
In terms of the Frobenius distance, the condition for the existence of an intersection point at $\tau$ reads $D_{\mathrm{F}}^{\rm I}(\tau)=D_{\mathrm{F}}^{\rm{II}}(\tau)$. Invoking the long-time approximation Eq. (\ref{rho_M}), this equality becomes
\begin{equation}\label{eq:s19}
    \mathrm{Tr}\left[(\hat{Z}^{\mathrm{I}}(t))^2\right]~=~\mathrm{Tr}\left[(\hat{Z}^{\mathrm{II}}(t))^2\right].
\end{equation}
Here, we have denoted $\hat{Z}(t)~=~\hat{M}(t)e^{-(\mu_{1}-\mu_{3})t}$ for simplicity. As $e^{-(\mu_{1}-\mu_{3})t}$ is just a function, Eq. (\ref{eq:s19}) reduces to
\begin{equation}
\mathrm{Tr}\left([\hat{M}^{\mathrm{I}}(\tau)]^2\right)~=~\mathrm{Tr}\left([\hat{M}^{\mathrm{II}}(\tau)]^2\right), \label{trace_FdA}
\end{equation}
The above equation is identical to the condition in Eq. (\ref{M_trace}). Hence the intersection time $\tau$ is still encoded in two solutions $[X(\tau)]_{\pm}$ given by Eq. (\ref{X_solution}) with $X(\tau)=e^{-(\mu_3-\mu_4)\tau}$.
To ensure the existence of a physically admissible intersection time $\tau$, we still impose the constraints Eqs. (\ref{X=1}) and (\ref{X_01}) on the solutions $[X(\tau)]_{\pm}$ in the regimes to the left and right of the Liouvillian exceptional point, respectively. Therefore, although the specific evolution of the Frobenius distance differs from that of the trace distance, the resulting conditions under which the intersection occurs are entirely consistent with those for the trace distance. 

We depict a set of numerical results in Figs. \ref{fig:full_Fd}. Overall, the results are akin to those showed in Fig. \ref{fig:long_trace_left}: (i) There are regimes with either single or multiple intersection points. (ii) The theoretical expression Eq. (\ref{X=1}) only captures the regime with multiple intersection points. Combining numerical results and theoretical analysis, we can ensure that the regime exhibiting a single intersection supports a genuine QMPE.

\subsection{Quantum relative entropy}
Another important measure that we consider for quantifying the existence of QMPE is the quantum relative entropy between the time-evolved state $\hat{\rho}(t)$ and the steady state $\hat{\rho}_{\mathrm{ss}}$, defined as
\begin{equation}
    S(t)=\mathrm{Tr}[\hat{\rho}(t)(\mathrm{ln}\hat{\rho}(t)-\mathrm{ln}\hat{\rho}_{\mathrm{ss}})].\label{S}
\end{equation}
$S(t)$ is a non-symmetric, information-theoretic quantity that measures how distinguishable the state $\hat{\rho}(t)$ is from the reference state $\hat{\rho}_{\mathrm{ss}}$. Combining the long time limit of $\hat{\rho}(t)$ in Eq. (\ref{rho_M}) with the approximation $\mathrm{ln}\left(\hat{\rho}_{\mathrm{ss}}+\hat{M}e^{-\left(\mu_{1}-\mu_{3}\right)t}\right
) ~\simeq~ \mathrm{ln}\left(\hat{\rho}_{\mathrm{ss}}\right)+e^{-\left(\mu_{1}-\mu_{3}\right)t}\hat{\rho}_{\mathrm{ss}}^{-1}\hat{M}$, we can approximate the quantum relative entropy in Eq. \eqref{S} as
\begin{equation}
     S(t)~\simeq~\mathrm{Tr}(\hat{M}^{2}\hat{\rho}_{\mathrm{ss}}^{-1})e^{-2(\mu_{1}-\mu_{3})t}.\label{S_sim}
\end{equation}
Here, the form of $\hat{M}$ is given below Eq. (\ref{rho_M}). This approximated form is amendable to analytical treatments.

Based on Eq. (\ref{S_sim}), the condition $S^{\rm I}(\tau)=S^{\rm{II}}(\tau)$ that signifies the occurrence of an intersection in terms of the quantum relative entropy reduces to
\begin{equation}
    \mathrm{Tr}((\hat{M}^{\mathrm{I}})^2\hat{\rho}_{\mathrm{ss}}^{-1})=\mathrm{Tr}((\hat{M}^{\mathrm{II}})^2\hat{\rho}_{\mathrm{ss}}^{-1}). \label{S_condition}
\end{equation}
Similarly, Eq. \eqref{S_condition} can be transferred into a quadratic equation for $\tilde{X}(\tau)=e^{-(\mu_3-\mu_4)\tau}$ with two solutions
\begin{equation}
[\tilde{X}(\tau)]_{\pm}=\frac{-2\tilde{P}(R_3^{\mathrm{I}}R_4^{\mathrm{I}}-R_3^{\mathrm{II}}R_4^{\mathrm{II}})\pm\sqrt{\tilde{\Delta}}}{2\tilde{T}_4 \left[ (R_4^{\mathrm{I}})^2-(R_4^{\mathrm{II}})^2 \right]},
\end{equation}
where $\tilde{\Delta}=4\tilde{P}^2(R_3^{\mathrm{I}}R_4^{\mathrm{I}}-R_3^{\mathrm{II}}R_4^{\mathrm{II}})^2-4\tilde{T}_3\tilde{T}_4 \left[(R_3^{\mathrm{I}})^2-(R_3^{\mathrm{II}})^2 \right] \left[(R_4^{\mathrm{I}})^2-(R_4^{\mathrm{II}})^2\right]$,  $\tilde{T}_3=\mathrm{Tr}[(\hat{\rho}_{3}-\hat{\rho}_{\mathrm{ss}})^2\hat{\rho}_{\mathrm{ss}}^{-1}]$, $\tilde{T}_4=\mathrm{Tr}[(\hat{\rho}_{4}-\hat{\rho}_{1})^{2}\hat{\rho}_{\mathrm{ss}}^{-1}]$, $\tilde{P}=\mathrm{Tr}[(\hat{\rho}_{3}-\hat{\rho}_{\mathrm{ss}})(\hat{\rho}_{4}-\hat{\rho}_{\mathrm{ss}})\hat{\rho}_{\mathrm{ss}}^{-1}]$. We remark that these solutions mirror the form of Eq. (\ref{X_solution}), differing only in the parameter substitutions $T_3 \rightarrow \tilde{T}_3$, $T_4\rightarrow \tilde{T}_4$, and $P \rightarrow \tilde{P}$. 

%-------------------------------------------------
\begin{figure}[t!] 
 \centering
\includegraphics[width=1\columnwidth]{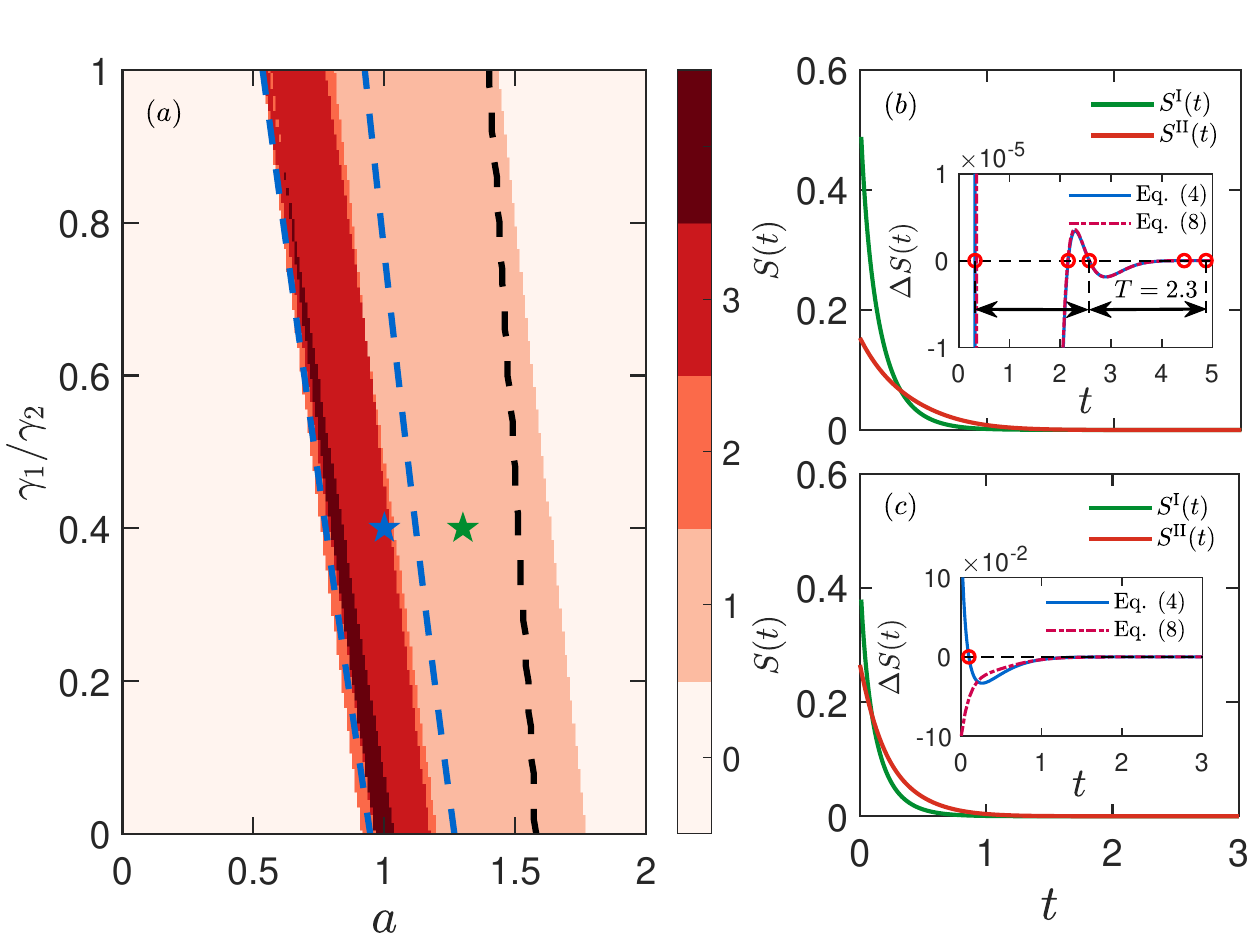} 
%\vspace*{-16mm}
\caption{(a) Contour plot of intersection point counts (non-negative integers) counted up to $t=20$ between two dynamical trajectories of $S(t)$ calculated using the full form of $\hat{\rho}(t)$ under the given two initial conditions. The blue dashed lines are analytical boundaries predicted by Eq. (\ref{X_tilde=1}). The black dashed line denotes the position of LEP. The blue ($a=1,\gamma_1=0.4$) and green ($a=1.3,\gamma_1=0.4$) stars mark the parameter sets used in (b) and (c) which depict dynamical trajectories of $S^{\rm{I,II}}(t)$. Insets in (b) and (c): Difference $\Delta S(t)=S^{\rm I}(t)-S^{\rm{II}}(t)$ calculated from Eq. (\ref{eq:rho_t}) (solid line) and Eq. (\ref{rho_M}) (dash-dotted line); red circles indicate exact intersection points. $T=2.3$ in inset of (b) denotes the theoretical prediction of period $\pi/|\mathrm{Im}[\mu_{3,4}]|$. $\gamma_2=1$ for all plots.}
\protect\label{fig:full_S}
\end{figure}
%-------------------------------------------------

Therefore, to provide analytical insights into the parameter regimes supporting the QMPE, we should also impose constraints on $[\tilde{X}]_{\pm}$, in analogy with those for the trace distance and the Frobenius distance. In the parameter regime to the left of the Liouvillian exceptional point, we impose the condition
\begin{equation}
    |[\tilde{X}(\tau)]_{\pm}|=1.\label{X_tilde=1}
\end{equation}
In the regime to the right of the Liouvillian exceptional point, we require that
\begin{equation}
    0<[\tilde{X}(\tau)]_{\pm}<1.
\end{equation}
Similar to the trace distance and the Frobenius distance described earlier, these constraints serve to ensure the presence of physical intersection points, thereby delineating the parameter regimes where the intersection actually occurs.

In Fig. \ref{fig:full_S}, we present numerical results for the number of intersections counted up to
$t=20$, along with representative dynamical trajectories. A comparison with Figs. \ref{fig:long_trace_left} and \ref{fig:full_Fd}--which show results for the trace distance and Frobenius distance, respectively--reveals that parameter regimes with either single or multiple intersections also exist here. Notably, the regime exhibiting a single intersection is broader than those identified using the trace distance or Frobenius distance. Interestingly, when compared with the theoretical regime bounded by blue dashed lines (which predicts an infinite number of intersections), we find that part of the numerically identified single-intersection regime falls within this theoretical region. This indicates that some intersections are missed in finite-time numerical counting. Consequently, we exclude this overlapping region as supporting the genuine QMPE. The genuine QMPE is thus identified only in regimes that exhibit a single intersection and lie outside the theoretically bounded area.

\section{Robustness of QMPE}\label{sec:3}
In the previous section, we identified parameter regimes supporting the genuine QMPE in a single PT-symmetric qubit system. Here, we examine the robustness of the QMPE against two factors: the ubiquitous dephasing noise present in real qubit systems, and an increase in the number of qubits.

\subsection{Dephasing effect}
%-------------------------------------------------
\begin{figure}[b!] 
 \centering
\includegraphics[width=1\columnwidth]{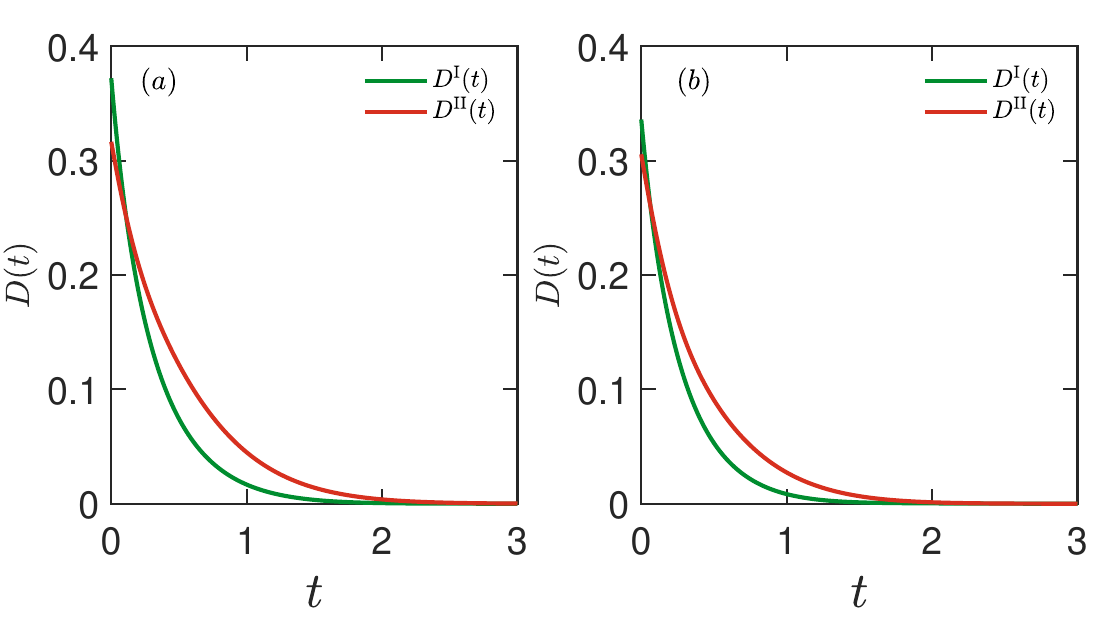} 
%\vspace*{-16mm}
\caption{Dynamical trajectories of the trace distance $D(t)$ under two initial states $\hat{\rho}^{\rm I}(0)=(\hat{\sigma}_z+\hat{\rm{I}})/2$ and $\hat{\rho}^{\rm{II}}(0)=\hat{\rm{I}}/2$. (a) $\gamma_3=0$. (b) $\gamma_3=0.2$. Other parameters are $a=1.3$, $\gamma_1=0.4$ and $\gamma_2=1$.} 
%notice: Eq.(2) without k=3
\protect\label{fig:dephasing}
\end{figure}
%-------------------------------------------------
We have known that dephasing typically accelerates the relaxation dynamics. In our previous model, however, we have only considered dissipation channels corresponding to the excitation and de-excitation processes of qubit induced by the bath. Since dephasing is almost unavoidable in realistic experiments, it is natural to ask whether the QMPE persists when the system is further subject to dephasing. In other words, we aim to examine whether the parameter region exhibiting the QMPE, as identified in the previous analysis, remains ``robust" against dephasing.

To describe the system under dephasing, in addition to the two dissipation channels defined in Eq. \eqref{rho(t)_Lindblad}, we further introduce a third pure dephasing channel. Altogether, the full Lindblad-type master equation takes the form
\bea
    \frac{d}{dt}\hat{\rho}(t) &=& -i\left[\hat{H}\hat{\rho}(t)-\hat{\rho}(t)\hat{H}^{\dagger}\right]+\sum_{k=1,2,3}\mathcal{D}_{k}[\hat{\rho}(t)]\nonumber\\
    &&-i\mathrm{Tr}[\hat{\rho}(t)(\hat{H}^{\dagger}-\hat{H})]\hat{\rho}(t),\label{rho(t)_Lindblad_dephasing}
\eea
where the additional term is characterized by the jump operator $\hat{L}_3=\hat{\sigma}_z$ and the dephasing strength $\gamma_3$. Figure \ref{fig:dephasing} shows the dynamical trajectories of the trace distance under the same parameter set for which the original PT-symmetric qubit system without dephasing exhibits the QMPE. While the detailed dynamics are modified by dephasing, the QMPE persists for moderate dephasing strengths, demonstrating a degree of robustness against such noise. This robustness significantly enhances the experimental feasibility of observing the QMPE in realistic 
PT-symmetric systems.

%-------------------------------------------------
\begin{figure*}[tbh!] 
 \centering
\includegraphics[width=2\columnwidth]{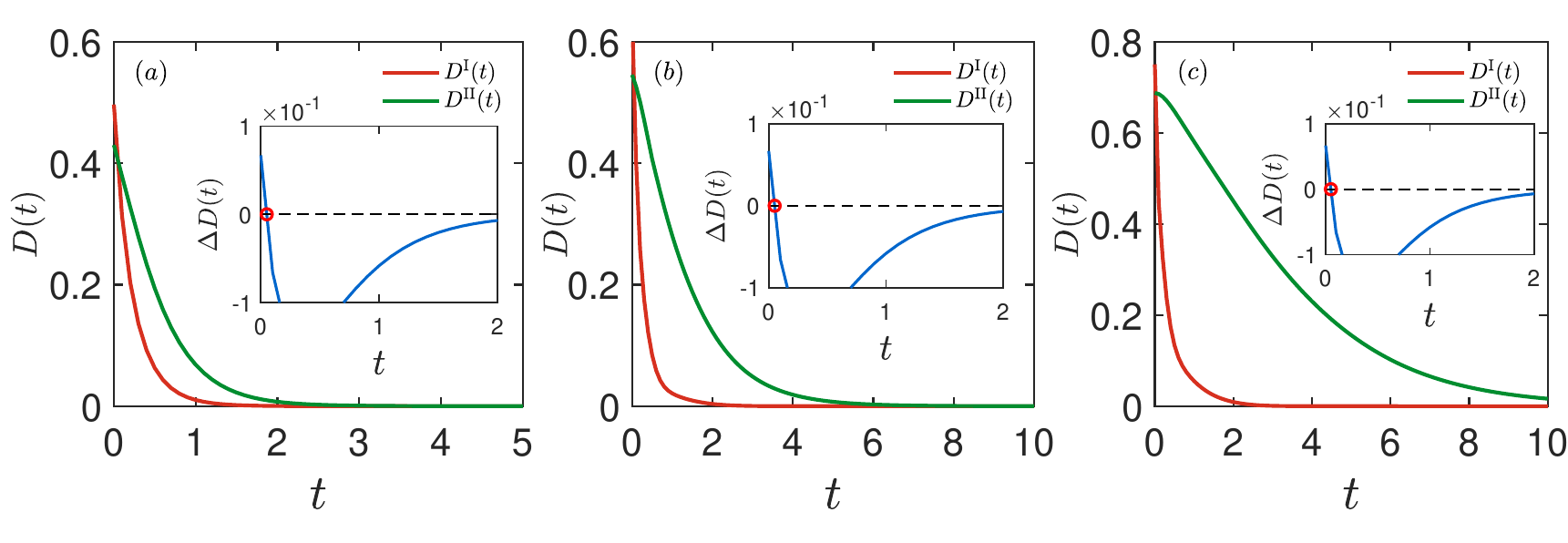} 
%\vspace*{-16mm}
\caption{The dynamical trajectories of $D^{\mathrm{I,II}}(t)$ starting from two specified initial conditions $(\hat{\rho}_s^{\mathrm{I}}(0))^{\otimes N}$ and $(\hat{\rho}_s^{\mathrm{II}}(0))^{\otimes N}$ for (a) $N=2$ and $a=1.2$, (b) $N=3, a=0.8$ and (c) $N=4, a=0.6$. Insets show the difference $\Delta D(t)=D^{\mathrm{I}}(t)-D^{\mathrm{II}}(t)$, with the red circles denoting the position of intersection point. Other parameters are $\gamma_1=0.1,\gamma_2=1$ for all plots.}
\protect\label{fig:Mbits_trace}
\end{figure*}
%-------------------------------------------------

\subsection{Multi-qubit systems}\label{a:3}
Now we examine whether the QMPE persists in PT-symmetric multiqubit systems. We consider a PT-symmetric multiqubit system through the tensor-product extension
\begin{equation}
    \hat{H}=\sum_{j=1}^{N}\hat{H}_{j,\mathrm{PT}}.
\end{equation}
Here, each single PT-symmetric qubit Hamiltonian $\hat{H}_{j,\mathrm{PT}}$ takes the form $\hat{\sigma}_x^j+ia\hat{\sigma}_z^j$ and $N\geqslant1$ marks the number of qubits which can be an arbitrary integer. This direct generalization ensures that the implementation of such multiqubit systems remains well within the reach of current experimental capabilities, particularly given recent
advances in the control of single and two PT-symmetric qubit systems. Moreover, such a design allows the composite systems to directly inherit the exact PT symmetry from their constituent building
blocks, thereby enabling well-defined and controllable
symmetry properties in more complex non-Hermitian multiqubit
systems. The dynamical evolution of the state $\hat{\rho}(t)$ of the composite system continues to adhere to Eq. \eqref{rho(t)_Lindblad}. And the role of the common bosonic thermal bath for the multiqubit system is captured through the jump operators $\hat{L}_1=\sum_{j=1}^{N}\hat{\sigma}_{+}^j$ and $\hat{L}_2=\sum_{j=1}^{N}\hat{\sigma}_{-}^j$, where the $\hat{\sigma}_{\pm}^j$ are spin ladder operator of $j$th qubit.

To demonstrate the occurrence of the QMPE, we explicitly consider two different initial states $\hat{\rho}^{\rm I}(0)=(\hat{\rho}_s^{\mathrm{I}}(0))^{\otimes N}$ and $\hat{\rho}^{\rm{II}}(0)=(\hat{\rho}_s^{\mathrm{II}}(0))^{\otimes N}$ which are tensor products between the same single qubit initial states $\hat{\rho}_s^{\mathrm{I,II}}(0)$, with $\hat{\rho}_s^{\rm I}(0)=\frac{1}{2}(\hat{\sigma}_z+\hat{\mathrm{I}})$ and $\hat{\rho}_s^{\rm{II}}(0)=\frac{1}{2}\hat{\mathrm{I}}$ ( ``$\hat{\mathrm{I}}$" is the identity matrix) that have been used in the single qubit scenario through the main text and the previous section. For simplicity, we here just focus on the trace distance $D(t)$ as the quantifier and depict dynamical trajectories of the trace distance $D^{\rm{I,II}}(t)$ that are sufficient to signify the occurrence of the QMPE in PT-symmetric multiqubit systems.

In Fig. \ref{fig:Mbits_trace}, we present a set of dynamical results for the trace distance $D^{\rm{I,II}}(t)$ under the given initial conditions for systems with two (a), three (b) and four (c) qubits. From the figure, we clearly observe the occurrence of QMPE as demonstrated by the existence of a single intersection point between two dynamical trajectories. This indicates that the QMPE is a generic phenomenon in PT-symmetric multiqubit systems.

\section{Discussion and conclusion}\label{sec:4}
In summary, we have demonstrated the occurrence of the QMPE in PT-symmetric qubit systems using established dynamical quantifiers, thereby extending the scope of this phenomenon beyond the well-studied settings of closed and open Hermitian systems. Crucially, the observed QMPE exhibits distinctive features that set it apart from previous manifestations: (i) it disappears immediately upon restoring Hermiticity in the Hamiltonian, and (ii) it arises independently of both Hamiltonian and Liouvillian exceptional points. These findings go beyond existing understandings based on Hermitian and effective non-Hermitian descriptions. 

Given the experimental feasibility of PT-symmetric qubit systems--already realized experimentally~\cite{Nori.22.PRR,Fang.21.CP}--our findings are readily amenable to experimental verification. Moreover, the QMPE persists in PT-symmetric multiqubit systems and under dephasing effect, underscoring its robustness. Our work not only deepens the fundamental understanding of the Mpemba phenomenon in the quantum realm, but also provides a crucial piece of the puzzle that confirms the ubiquity of the QMPE across both Hermitian and non-Hermitian regimes.

\section*{Acknowledgments}
J.L. acknowledges support from the National Natural Science Foundation of China (Grant No. 12205179) and start-up funding of Shanghai University.

%%====================== appendices ================
%\clearpage
%\onecolumngrid
\appendix

\renewcommand{\theequation}{A\arabic{equation}}
\setcounter{equation}{0}  % reset counter
\section{Complete suppression of the slowest relaxation mode} \label{a:1}
In this section, we discuss properties of the expansion coefficients $C_j$ of $\hat{\rho}(t)$ in the eigenbasis of the Liouvillian superoperator $\mathcal{L}_0$. Based on the quantum master equation Eq. (2) in the main text, the matrix form of $\mathcal{L}_0$ for a PT-symmetric qubit model reads
\begin{equation}
\mathcal{L}_0=\begin{bmatrix}
2a-\gamma_2&-i&i&\gamma_1\\-i&-0.5\gamma_1-0.5\gamma_2&0&i\\i&0&-0.5\gamma_1-0.5\gamma_2&-i\\\gamma_2&i&-i&-2a-\gamma_1
\end{bmatrix}.\label{L0}\end{equation}
The spectrum of $\mathcal{L}_0$ is introduced through the relations $\mathcal{L}_0\hat{\rho}_j=\mu_j\hat{\rho}_j$ and $\mathcal{L}_{0}^{\dagger}\hat{\omega}_{j}=\mu_{j}^*\hat{\omega}_{j}$,where $\hat{\rho}_j$ and $\hat{\omega}_j$ are the right and left normalized eigenvectors of $\mathcal{L}_0$ with an eigenvalue $\mu_j$ and its complex conjugate $\mu_j^{\ast}$, respectively. Denoting $\operatorname{vec}(X)=(X_{11},X_{21},X_{12},X_{22})^T$ as the vectorized form of a matrix $X=\begin{bmatrix}X_{11}&X_{12}\\X_{21}&X_{22}\end{bmatrix}$ with $T$ the transpose operation, we introduce the swap matrix 
\begin{equation}
S~=~\begin{bmatrix}1&0&0&0\\0&0&1&0\\0&1&0&0\\0&0&0&1\end{bmatrix},
\end{equation}
which can swap the second and third components of $\operatorname{vec}(X)$ such that $\operatorname{vec}(X^\dagger)=(X_{11}^{\ast},X_{12}^{\ast},X_{21}^{\ast},X_{22}^{\ast})^{T}=S\operatorname{vec}(X)^{\ast}$. We readily observe that 
\begin{equation}
    \mathcal{L}_0=S\mathcal{L}_0^{\ast}S,\label{HP}
\end{equation}
since parameters $a$ and $\gamma_{1,2}$ are all real. This property Eq. (\ref{HP}) implies that $\mathcal{L}_0$ is Hermitian-preserving, meaning that for any matrix $X$ we have $ (\mathcal{L}_0X)^\dagger=\mathcal{L}_0(X^\dagger)$. Therefore, $\hat{\rho}_1$ which defines the steady state $\hat{\rho}_{\mathrm{ss}}$ is naturally Hermitian with the property $\hat{\omega}_1^{\dagger}=\hat{\rho}_1$. We can thus infer that $C_1=\mathrm{Tr}[\hat{\omega}_1^\dagger\hat{\rho}(0)]$ is real.

In addition, utilizing the matrix form in Eq. \eqref{L0}, we can express the equation $\mathcal{L}_0\hat{\rho}_j=\mu_j\hat{\rho}_j$ as 
\begin{widetext}
\begin{equation}
\begin{bmatrix}
2a-\gamma_2 & -i & i & \gamma_1 \\
-i & -0.5\gamma_1-0.5\gamma_2 & 0 & i \\
i & 0 & -0.5\gamma_1-0.5\gamma_2 & -i \\
\gamma_2 & i & -i & -2a-\gamma_1
\end{bmatrix}
\begin{bmatrix}
\rho_{j,11} \\
\rho_{j,21} \\
\rho_{j,12} \\
\rho_{j,22}
\end{bmatrix}\\
=\mu_j\begin{bmatrix}
\rho_{j,11} \\
\rho_{j,21} \\
\rho_{j,12} \\
\rho_{j,22}
\end{bmatrix}.
\label{eigenfunction}
\end{equation}
\end{widetext}
Here, $\{\rho_{j,mn}\}$ denote elements of the vectorized form of $\hat{\rho}_j$ according to our convention. To proceed, we write down explicit equations for each element
\begin{equation}
(2a-\gamma_2)\rho_{j,11}-i\rho_{j,21}+i\rho_{j,12}+\gamma_1\rho_{j,22}=\mu_j\rho_{j,11},\label{eigeneq_1}
\end{equation}
\begin{equation}
-i\rho_{j,11}-0.5(\gamma_1+\gamma_2)\rho_{j,21}+i\rho_{j,22}=\mu_j\rho_{j,21},\label{eigeneq_2}
\end{equation}
\begin{equation}
i\rho_{j,11}-0.5(\gamma_1+\gamma_2)\rho_{j,12}-i\rho_{j,22}=\mu_j\rho_{j,12},\label{eigeneq_3}
\end{equation}
\begin{equation}
    \gamma_2\rho_{j,11}+i\rho_{j,21}-i\rho_{j,12}-(2a+\gamma_1)\rho_{j,22}=\mu_j\rho_{j,22}.\label{eigeneq_4}
\end{equation}

Summing up Eqs. \eqref{eigeneq_2} and \eqref{eigeneq_3}, we can obtain the following relation between the eigenvalues and the off-diagonal elements of the matrix $\hat{\rho}_j$
\begin{equation}
[\mu_j+0.5(\gamma_1+\gamma_2)](\rho_{j,12}+\rho_{j,21})=0.\label{eigeneq_offdia}
\end{equation}
Similarly, Eqs. \eqref{eigeneq_1} and \eqref{eigeneq_4} together yield a relation between the eigenvalues and the diagonal elements of right eigenmatrix $\hat{\rho}_j$
\begin{equation}
    (2a-\mu_j)\rho_{j,11}=(2a+\mu_j)\rho_{j,22}.\label{eigeneq_dia}
\end{equation}

For PT-symmetric qubit system considered here, we have $\mu_2=-0.5(\gamma_1+\gamma_2)$ and $\mu_j\neq-0.5(\gamma_1+\gamma_2)$ for $j\neq 2$. According to Eq. \eqref{eigeneq_2} or \eqref{eigeneq_3}, we first find $\rho_{2,11}=\rho_{2,22}$. Inserting this relation into Eq. \eqref{eigeneq_dia}, we can only infer that
 \begin{equation}
    \rho_{2,11}=\rho_{2,22}=0, \label{rho_2diag}
\end{equation}
since $\mu_2$ and $a$ are nonzero. The above relation implies that $\hat{\rho}_2$ is a traceless matrix. By substituting Eq. \eqref{rho_2diag} into Eq. \eqref{eigeneq_1} or \eqref{eigeneq_4}
, we obtain the relation between two off-diagonal elements of $\hat{\rho}_2$
 \begin{equation}
    \rho_{2,21}=\rho_{2,12}. \label{rho_2offdiag}
\end{equation}
One can easily check that the relations in Eqs. (\ref{rho_2diag}) and (\ref{rho_2offdiag}) extend to the left eigenmatrix, yielding $\omega_{2,11}=\omega_{2,22}=0, \omega_{2,21}=\omega_{2,12}$. Using the above relations among elements of $\hat{\omega}_2$, we find
\begin{equation}
 C_2~=~\mathrm{Tr}[\hat{\omega}_2^\dagger\hat{\rho}(0)]~=~\omega_{2,12}[\rho_{12}(0)+\rho_{21}(0)].
\end{equation}
Hence, for any incoherent (i.e., diagonal) initial states or coherent initial ones with purely imaginary off-diagonal elements, we always have $C_2=0$.  

We now analyze the expansion coefficients $C_{3,4}$. In the regime left to the Liouvillian exceptional point, we have $\mu_3=\mu_4^{\ast}$ forming a complex conjugate pair. We note that the Hermitian-preserving condition [Eq. \eqref{HP}] implies
\begin{equation}
   \mathcal{L}_0 (S \hat{\rho}^{\ast}_j)=\mu^{\ast}_j (S \hat{\rho}^{\ast}_j),
\end{equation}
namely, if ($\mu_{j},\hat{\rho}_j$) is an eigenpair of $\mathcal{L}_0$, then ($\mu_{j}^{\ast},\hat{\rho}_{j}^{\dagger}$) is also an eigenpair. Then $\mu_3=\mu_4^{\ast}$ implies $\hat{\rho}_3=\hat{\rho}_4^{\dagger}$, and consequently,
\begin{equation}
\begin{gathered}
    C_4=\mathrm{Tr}[\hat{\omega}_4^\dagger\hat{\rho}(0)]=\mathrm{Tr}[\hat{\omega}_3\hat{\rho}(0)]=C_3^{\ast},
\end{gathered}
\end{equation}
where the last relation follows from $\mathrm{Tr}(A^\dagger B)=[\mathrm{Tr}(B^\dagger A)]^{\ast}$. In the regime right to the Liouvillian exceptional point, there is no relation between $C_{3,4}$ and we just have $\mu_{3,4}\in \mathbb{R}$ and $C_{3,4}\in \mathbb{R}$.

\renewcommand{\theequation}{B\arabic{equation}}
\setcounter{equation}{0}  % reset counter
\section{Long time approximation of $\hat{\rho}(t)$} \label{a:2}
In this section, we show how to get a long time approximation of $\hat{\rho}(t)$ [Eq. (7) in the main text]. We first note that the decomposition of $\hat{\rho}(t)$ in the eigenbasis of $\mathcal{L}_0$ can be rewritten as
\begin{equation}\label{eq:s16}
\begin{aligned}
    \hat{\rho}(t)=&\frac{\sum_je^{\mu_jt}C_j\hat{\rho}_j}{\mathrm{Tr}\left(\sum_je^{\mu_jt}C_j\hat{\rho}_j\right)}\\
    =&\frac{\hat{\rho}_1+\sum_{j\geqslant2}\frac{C_j}{C_1}e^{-(\mu_1-\mu_j)t}\hat{\rho}_j}{1+\sum_{j\geqslant2}\frac{C_j}{C_1}e^{-(\mu_1-\mu_j)t}}.
\end{aligned}
\end{equation}
We then invoke the approximation $\frac{1}{1+x}\approx1-x$ valid under $\left|x\right|\ll1$ and simplify Eq. (\ref{eq:s16}) as
\begin{equation}\label{eq:s17}
    \begin{aligned}
\hat{\rho}(t)
~\simeq~&\left( \hat{\rho}_{1}+\sum_{j\geq2}\frac{C_{j}}{C_{1}}e^{-\left(\mu_{1}-\mu_{j}\right)t}\hat{\rho}_{j}\right) \left(1-\sum_{j\geq2}\frac{C_{j}}{C_{1}}e^{-\left(\mu_{1}-\mu_{j}\right)t}\right)\\
=&\hat{\rho}_{1}+\sum_{j\geq2}\frac{C_{j}}{C_{1}}e^{-\left(\mu_{1}-\mu_{j}\right)t}\left(\hat{\rho}_{j}-\hat{\rho}_{1}\right)+\mathcal{O}(e^{-\Delta_{l,k}t})\\
~\simeq~&\hat{\rho}_{1}+\sum_{j\geq2}\frac{C_{j}}{C_{1}}e^{-\left(\mu_{1}-\mu_{j}\right)t}\left(\hat{\rho}_{j}-\hat{\rho}_{1}\right),
\end{aligned}
\end{equation}
where $\mathcal{O}(e^{-\Delta_{l,k}t})$ with $\Delta_{l,k}=2\mu_1-\mu_{l}-\mu_{k} (l,k >1)$ represent higher-order correction terms that decay exponentially as $e^{-\Delta_{l,k}t}$ and become negligible in the long time limit. The last form in Eq. (\ref{eq:s17}) corresponds to Eq. (7) of the main text by noting that $\hat{\rho}_1=\hat{\rho}_{\rm{ss}}$ and $C_2=0$ under the chosen initial states.

%\bibliography{NHM}
\end{document}